\documentclass[12pt]{article}

\usepackage{acronym}
\usepackage{amsmath,amssymb}
\usepackage[backend=biber,style=nature,citestyle=nature]{biblatex}
\usepackage{bm}
\usepackage{booktabs}
\usepackage[format=plain,labelformat=simple,font=small]{caption}
\usepackage[final]{changes}
\usepackage{color}
\usepackage{fullpage}
\usepackage[top=2cm, bottom=2cm, left=2cm, right=2cm]{geometry} 
\usepackage{graphicx}
\usepackage{nameref}
\usepackage[hidelinks]{hyperref}
    \usepackage{cleveref} 
\usepackage[utf8]{inputenc}
\usepackage[pagewise]{lineno}
\usepackage{microtype}
\usepackage{multirow}
\usepackage{placeins}
\usepackage{pslatex} 
\usepackage{sectsty}
\usepackage{soul}
\usepackage{tabularx}       
\usepackage{todonotes}

\definechangesauthor[color=blue]{JJ}
\definechangesauthor[color=red]{WS}
\definechangesauthor[color=brown]{WW}

\newcommand{\appropto}{\mathrel{\vcenter{
  \offinterlineskip\halign{\hfil$##$\cr
    \propto\cr\noalign{\kern2pt}\sim\cr\noalign{\kern-2pt}}}}}

\title{\replaced[id=JJ]{Conductance-based dendrites perform Bayes-optimal cue integration}{Learning Bayes-optimal dendritic opinion pooling}}

\author{Jakob Jordan$^{1}$\footnote{Correspondence:~jakob.jordan@unibe.ch}, Jo\~{a}o Sacramento$^{1,3}$, Willem A.M. Wybo$^{1,4}$,\\Mihai A. Petrovici$^{1}$\footnote{Joint senior authorship.} \,\& Walter Senn$^{1\dagger}$ \\[1em]
  \small $^1$Department of Physiology, University of Bern, Bern, Switzerland\\
  \small $^3$Institute of Neuroinformatics, UZH / ETH Zurich, Zurich, Switzerland\\
  \small $^4$Institute of Neuroscience and Medicine, Forschungszentrum J\"ulich, J\"ulich, Germany
}

\date{\small \today}

\newcommand{\alphasd}{\alpha^{\text{sd}}}
\newcommand{\alphads}{\alpha^{\text{ds}}}
\newcommand{\bargs}{\bar{g}_\text{s}}

\newcommand{\dotus}{\dot{u}_\text{s}}
\newcommand{\E}{\text{E}}
\newcommand{\Ed}{E^\text{d}}
\newcommand{\vecEd}{\boldsymbol{E}^\text{d}}
\newcommand{\EE}{E^\text{E}}
\newcommand{\EI}{E^\text{I}}
\newcommand{\EL}{E^\text{L}}
\newcommand{\Eprior}{E_0}
\newcommand{\barEs}{\bar{E}_\text{s}}
\newcommand{\gd}{g^\text{d}}
\newcommand{\vecgd}{\boldsymbol{g}^\text{d}}
\newcommand{\gE}{g^\text{E}}
\newcommand{\gI}{g^\text{I}}
\newcommand{\gL}{g^\text{L}}
\newcommand{\gdsc}{g^{\text{ds}}}
\newcommand{\gsdc}{g^{\text{sd}}}
\newcommand{\gprior}{g_0}
\newcommand{\I}{\text{I}}
\newcommand{\Lk}{\text{L}}
\newcommand{\lambdae}{\lambda_\text{e}}
\newcommand{\rs}{r_\text{s}}

\newcommand{\us}{u_\text{s}}
\newcommand{\Zprior}{Z_0}
\newcommand{\Zd}{Z^\text{d}}
\newcommand{\ud}{u^\text{d}}
\newcommand{\tildeEd}{\tilde{E}^\text{d}}
\newcommand{\Cd}{C^\text{d}}

\newcommand{\wE}{w^\text{E}}
\newcommand{\wI}{w^\text{I}}

\newcommand{\mV}{\,\text{mV}}
\newcommand{\ms}{\,\text{ms}}
\newcommand{\s}{\,\text{s}}
\newcommand{\pF}{\,\text{pF}}
\newcommand{\pers}{\,\frac{1}{\text{s}}}
\newcommand{\nS}{\,\text{nS}}
\newcommand{\mydeg}{\,\text{deg}}

\newcommand{\todoinhide}[1]{}

\begin{document}

\maketitle

\begin{abstract}  
  \replaced[id=JJ]{A fundamental function of cortical circuits is the integration of information from different sources to form a reliable basis for behavior.
  While animals behave as if they optimally integrate information according to Bayesian probability theory, the implementation of the required computations in the biological substrate remains unclear.
  We propose a novel, Bayesian view on the dynamics of conductance-based neurons and synapses which suggests that they are naturally equipped to optimally perform information integration.
  In our approach apical dendrites represent prior expectations over somatic potentials, while basal dendrites represent likelihoods of somatic potentials.
  These are parametrized by local quantities, the effective reversal potentials and membrane conductances.
  We formally demonstrate that under these assumptions the somatic compartment naturally computes the corresponding posterior.
  }{In functional network models, neurons are commonly conceptualized as linearly summing presynaptic inputs before applying a non-linear gain function to produce output activity.
  In contrast, synaptic coupling between neurons in the central nervous system is regulated by dynamic permeabilities of ion channels.
  So far, the computational role of these membrane conductances remains unclear and is often considered an artifact of the biological substrate.
  Here we demonstrate that conductance-based synaptic coupling allow neurons to represent, process and learn uncertainties.
  We suggest that membrane potentials and conductances on dendritic branches code opinions with associated reliabilities.
  The biophysics of the membrane combines these opinions by taking account their reliabilities, and the soma thus acts as a decision maker.}
  We derive a gradient-based plasticity rule, allowing neurons to learn desired target distributions and weight synaptic inputs by their relative reliabilities.
  Our theory explains various experimental findings on the system and single-cell level related to multi-sensory integration,\added[id=JJ]{ which we illustrate with simulations.
  Furthermore}, we make \added[id=JJ]{experimentally }testable predictions on Bayesian dendritic integration and synaptic plasticity.
\end{abstract}


\section*{Introduction}

\replaced[id=JJ]{Successful actions are based on information gathered from a variety of sources.}{Successful decision making is based on well-considered arguments.}
This holds as true for individuals as it does for whole societies.
For instance, \replaced[id=JJ]{experts, political parties, and special interest groups may all have different opinions on proposed legislature.}{opinions on proposed legislature may vary between experts, political parties and special interest groups.}
How should one combine these different \replaced[id=JJ]{views}{opinions}?
One might, for example, \replaced[id=JJ]{weight}{integrate the different opinions by weighting} them according to their relative reliability, estimated from \deleted[id=JJ]{their past performance, or} demonstrated expertise.
\replaced[id=JJ]{According to Bayesian probability theory, the combined reliability-weighted view contains more information than any of the individual views taken on its own and thus provides an improved basis for subsequent actions \cite{jaynes2003probability}}{The final decision can then be based on the joint, reliability-weighted opinion, representing a compromise}.

Such problems of weighting and combining \replaced[id=JJ]{information from different sources}{different opinions} are commonplace for our brains.
Whether inputs from neurons with different receptive fields or inputs from different modalities (Fig.~\ref{fig:intro-bayes-optimal}a), our cortex needs to combine these uncertain information sources into a coherent \replaced[id=JJ]{basis that enables informed actions}{whole}.
\added[id=JJ]{Bayesian probability theory provides clear recipes for how to optimally solve such problems, but so far the implementation in the biological substrate is unclear.}
Previous work has demonstrated that multiple interacting neuronal populations can efficiently perform such probabilistic computations \cite{ma2006bayesian,echeveste2020cortical}.
These studies provided mechanistic models potentially underlying the often Bayes-optimal behavior observed in humans and other animals \cite{ernst2002humans,knill2003humans,hillis2004slant}.
Here we demonstrate that probabilistic computations \replaced[id=JJ]{may be}{are} even deeper ingrained in our biological substrate, in single cortical neurons.

\added[id=JJ]{We suggest that each dendritic compartment, here interpreted as logical subdivision of a complex morphology, represents either a (Gaussian) likelihood function or a (Gaussian) prior distribution over somatic potentials.
These are parametrized by the local effective reversal potential and the membrane conductance.
Basal dendrites receiving bottom-up input represent likelihoods, while apical dendrites receiving top-down input, represent priors.}
\deleted[id=JJ]{In cortical neurons, each dendritic branch receives information from presynaptic partners and forms a local membrane potential.
We propose this to be the analog of an opinion.
In the absence of other compartments and leak currents, the somatic output, the analog of a decision, would reflect the opinion of the single branch.
However, in the presence of the leak and multiple branches, the soma encodes a reliability-weighted combination of a prior and additional opinions.
We further propose that the reliability of a dendritic branch with regard to a particular local opinion is encoded in its local conductance, including conductances elicited by synaptic input.
The biophysics of the bidirectional current flow in cortical neurons with multiple dendritic compartments naturally implements Bayesian opinion weighting (Fig.~\ref{fig:intro-bayes-optimal}b), while the output of the neuron encodes decisions based on the pooled opinions.}
\replaced[id=JJ]{We show that the natural dynamics of leaky integrator models compute the corresponding posterior}{Formally, the neuronal operation can be described as computing a posterior distribution}.
The crucial ingredient is the divisive normalization \replaced[id=JJ]{of compartmental membrane potentials naturally performed in the presence of}{performed by} conductance-based synaptic coupling \cite{carandini1994summation}.
\replaced[id=JJ]{Furthermore, while this computation relies on bidirectional coupling between neuronal compartments}{While the dendritic opinion weighting emerges from the recurrent interaction of multiple compartments within the dendritic tree}, at the level of the neuronal input-output transfer function, the effective computation can be described in a feed-forward manner.

Beyond \replaced[id=JJ]{performing inference}{opinion weighting itself}, the single-neuron view \added[id=JJ]{of reliability-weighted integration }provides an efficient basis for learning.
\replaced[id=JJ]{In our approach, s}{S}ynapses not only learn to reproduce a somatic target activity \cite{urbanczik2014learning}, but they also adjust synaptic weights to achieve some target variance in the somatic potential.
Furthermore, afferents with low reliability will be adjusted to contribute with a smaller total excitatory and inhibitory conductance to allow other projections to gain more influence.
Implicitly, this allows each dendritic compartment to adjust its relative reliability according to its past success in contributing to \replaced[id=JJ]{matching desired somatic distributions}{decisions}.

In our theoretical framework we derive somatic membrane potential dynamics and synaptic plasticity jointly via stochastic gradient ascent on the log-posterior distribution of somatic potentials.
Simulations demonstrate successful learning of a prototypical \replaced[id=JJ]{multisensory integration task}{opinion weighting task, and the integration of sensory cues from different modalities to guide behavior}.
The trained model allows us to interpret behavioral and neuronal data from cue integration experiments through a \replaced[id=JJ]{Bayesian}{computational} lens and to make specific predictions about both system behavior and single cell dynamics.

\section*{Results}

\subsection*{\replaced[id=JJ]{Integration of uncertain information}{Opinion weighting} in cortical neurons}

\begin{figure}[tbp!]
    \centering
    \includegraphics[width=1.\textwidth]{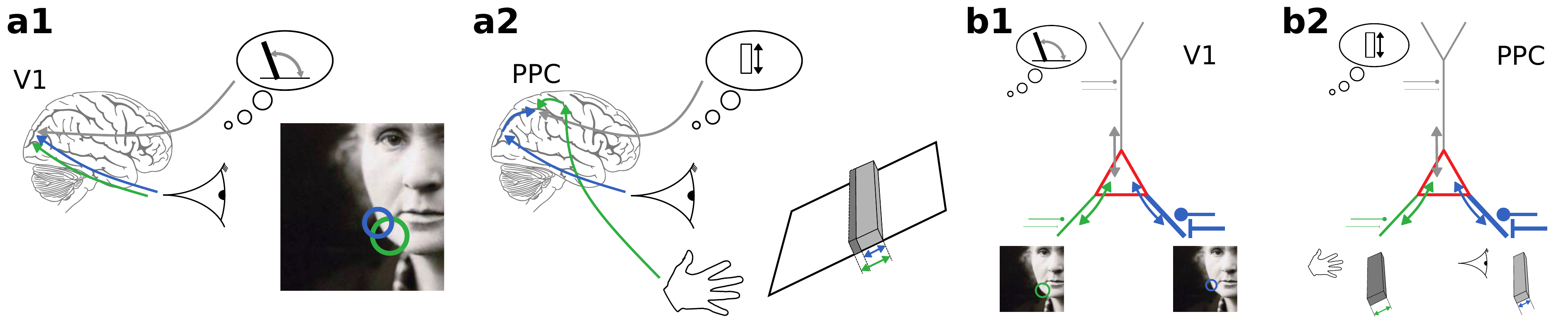}
    \caption{
      {\bf \replaced[id=JJ]{Integration of uncertain information in cortical neurons}{Dendritic opinion weighting as a canonical neuronal operation across cortex}.}
      {\bf (a1)} Cue integration in early visual \replaced[id=JJ]{processing}{perception} judging the orientation of a local edge.
      {\bf (a2)} Cue integration in multimodal \replaced[id=JJ]{perception}{processing} judging the height of a bar~\cite{ernst2002humans}.
      {\bf (b1)} \replaced[id=JJ]{A neuron integrates visual cues and prior expectations to combine}{Dendritic opinion weighting of visual cues combining} information across receptive fields.
      {\bf (b2)} \replaced[id=JJ]{A neuron integrates visual and haptic cues with prior expectations to combine}{Dendritic opinion weighting of multisensory cues combining} information across modalities.
      \replaced[id=JJ]{These computations can be}{This probabilistic computation is} realized by \replaced[id=JJ]{the natural dynamics of cortical neurons through the bidirectional coupling of compartments (colored arrows) which represent likelihood functions (green, blue), prior (grey), or posterior distributions (red) through their local membrane conductance and effective reversal potential}{the bidirectional voltage propagation in cortical neurons (colored arrows) that settles at the pooled somatic opinion (red triangle).
      The somatic potential represents the reliability-weighted dendritic opinions (grey, green, blue), calculated by a biophysical "consensus finding".}
    }\label{fig:intro-bayes-optimal}
\end{figure}

\replaced[id=JJ]{To give a high-level intuition for our approach, let us consider a prototypical task our brains have to solve}{We consider a prototypical example of neuronal opinion weighting}: the integration of various cues about a stimulus, for example in early visual areas from different parts of the visual field (Fig.~\ref{fig:intro-bayes-optimal}a) or in association areas from different sensory modalities (Fig.~\ref{fig:intro-bayes-optimal}b).

Due to properties of the stimulus and of our sensory systems, information delivered via various modalities inherently differs in reliability.
Behavioral evidence demonstrates that humans and non-human animals are able to integrate sensory input from different modalities \cite[e.g.,][]{rock1964vision,ernst2002humans,knill2003humans,hillis2004slant,alais2004ventriloquist,fetsch2009dynamic,fischer2011owl,raposo2012multisensory,nikbakht2018supralinear} and prior experience \cite[e.g.,][]{xu2017adaptive,darlington2018neural}, to achieve a similar performance as Bayes-optimal cue-integration models.
\replaced[id=JJ]{Our theory suggests that pyramidal cells are naturally suited to implement the necessary computations. In particular they take both their inputs and their respective reliabilities into account by}{We suggest that pyramidal cells across cortex naturally take the average reliability of their inputs into account} using two orthogonal information channels: membrane potentials and conductances.

\begin{figure}[tbp]
    \centering
    \includegraphics[width=.9\textwidth]{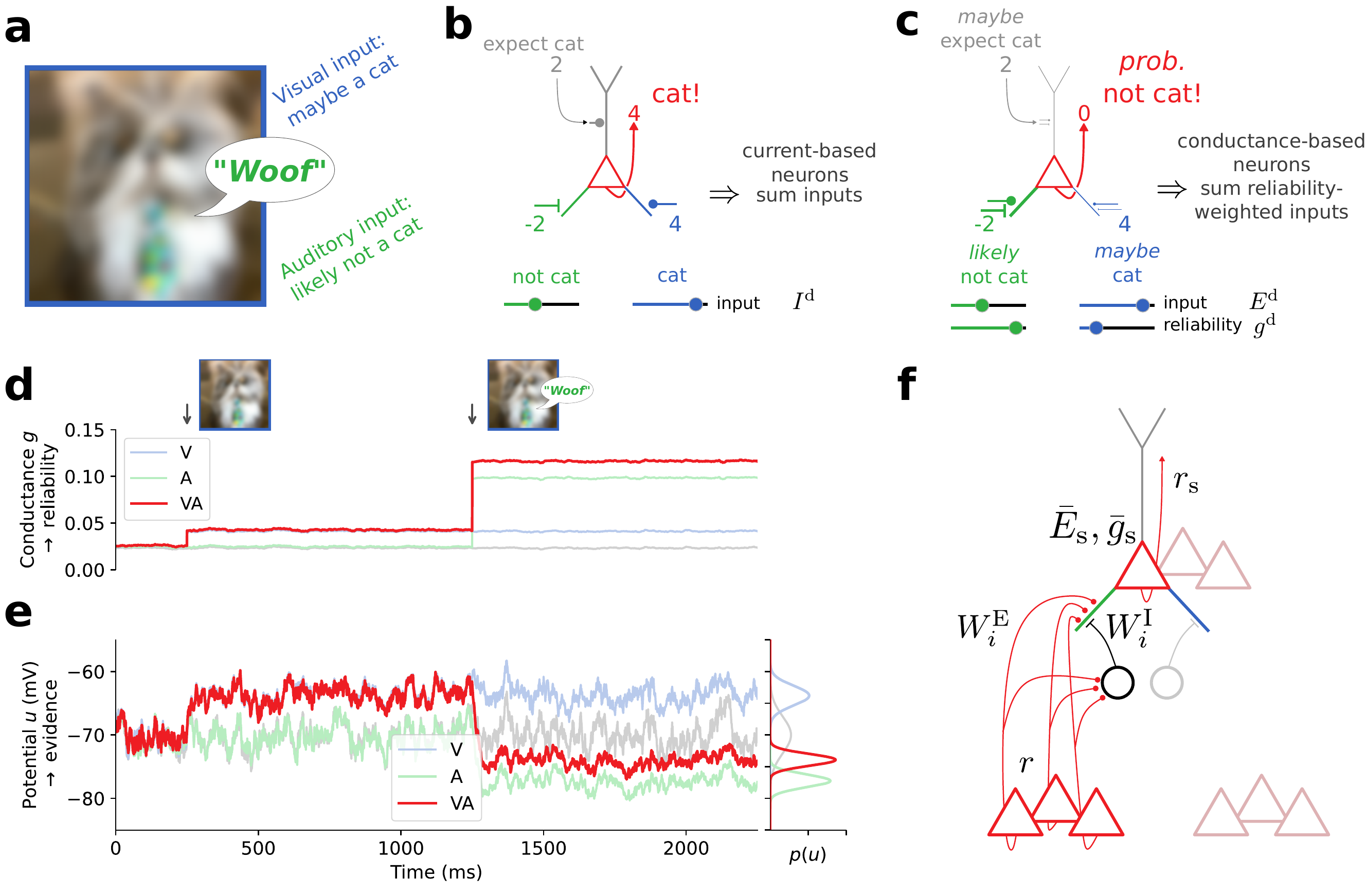}
    \caption{
        {\bf Conductance-based neuronal dynamics naturally implement \replaced[id=JJ]{Bayesian}{probabilistic} cue integration.}
        {\bf (a)} A multisensory stimulus.
        {\bf (b)} Current-based neuron\replaced[id=JJ]{ models}{s} can only additively accumulate \replaced[id=JJ]{information about}{opinions} their preferred feature.
        {\bf (c)} Conductance-based neuron models simultaneously represent \replaced[id=JJ]{information}{opinions} and associated reliability.
        {\bf (d)} Total somatic conductances $\bargs$ consisting of leak and synaptic conductances in a multisensory neuron \added[id=JJ]{(see panel (c)) }under three conditions: only visual input (V, blue), only auditory input (A, green), bimodal input (VA, red), and no input (gray).
        Before $400\text{ms}$ the visual cue is absent.
        Before $1200\text{ms}$ the auditory cue is absent.
        {\bf (e)} Somatic membrane potentials $\us$ are noisy, time-continuous processes that sample from the somatic distributions in the respective condition.
        This histogram on the right shows the somatic potential distributions between $1250\text{ms}$ and $2250\text{ms}$.
        {\bf (f)} Suggested microcircuit implementation.
        \added[id=JJ]{Top part shows the neuron from panel (c).}
        Activity $r$ of pyramidal cells from lower areas is projected directly (red lines with circular markers, $W^\E_i$ denote excitatory synaptic weights) and indirectly via inhibitory interneurons (circles and black lines with bar markers, $W^\I_i$ denote inhibitory synaptic weights) to different dendritic compartments of pyramidal cells in higher cortical areas.
        Each pyramidal cell represents \replaced[id=JJ]{pooled information}{a pooled opinion} $\barEs$ with \replaced[id=JJ]{its associated reliability}{some associated inverse variance} $\bargs$ distributed across a corresponding population (overlapping triangle triples, representing pre- and postsynaptic \replaced[id=JJ]{neurons}{opinions}, respectively).
        }\label{fig:results-cond-neuron-sketch}
\end{figure}

Consider a situation where your visual sensory apparatus is impaired, for example, due to a deformation of the lens.
Presented with multimodal stimuli that provide auditory and visual cues, you would have learned to rely more on auditory cues rather than visual input (Fig.~\ref{fig:results-cond-neuron-sketch}).
When confronted with an animal as in Fig.~\ref{fig:results-cond-neuron-sketch}a, based on your vision alone, you might expect it to be a cat, but not be certain about it.
Hearing it bark, however, would shift your belief towards it being, with high certainty, a dog.
Since current-based neuron models only encode \replaced[id=JJ]{information}{opinions} about their preferred feature in the total synaptic current without considering the relative reliability of different pathways, they can generate wrong decisions: here, a neuron that integrates auditory and visual cues wrongly signals the presence of a cat to higher cortical areas (Fig.~\ref{fig:results-cond-neuron-sketch}b).
In contrast\added[id=JJ]{, as we will show in the next section}, by using dendritic conductances $\gd$ as an additional coding dimension besides effective dendritic reversal potentials $\Ed$, conductance-based neuron models are able to respond correctly by weighting auditory inputs stronger than visual inputs (Fig.~\ref{fig:results-cond-neuron-sketch}c).
\replaced[id=JJ]{Intuitively, i}{I}n the absence of stimuli, the ``cat neuron'' \replaced[id=JJ]{(Fig.~\ref{fig:results-cond-neuron-sketch}b,c) represents a small (prior) probability}{has a low prior opinion} that a cat may be present, \replaced[id=JJ]{and}{but clearly increases this opinion upon} the presentation of an ambiguous cat-dog image \added[id=JJ]{increases this probability }(Fig.~\ref{fig:results-cond-neuron-sketch}e, $400-1200\text{ms}$, d,e).
\replaced[id=JJ]{However, w}{W}hen the animal subsequently barks, the \replaced[id=JJ]{probability}{opinion about the presence of a cat} drops\replaced[id=JJ]{abruptly. In our approach these computations are reflected by a hyperpolarization of the somatic membrane potential and an associated increase in membrane conductance}{, i.e., the somatic membrane potential of the cat neuron hyperpolarizes, while the reliability for this updated opinion increases, i.e., conductances increase.}
Consistent with Bayes-optimal cue-integration models \cite[e.g.,][]{knill2004bayesian}, the combined estimate shows an increased reliability, even if the cues are opposing.

\begin{figure}[t!]
    \centering
    \includegraphics[width=1.\textwidth]{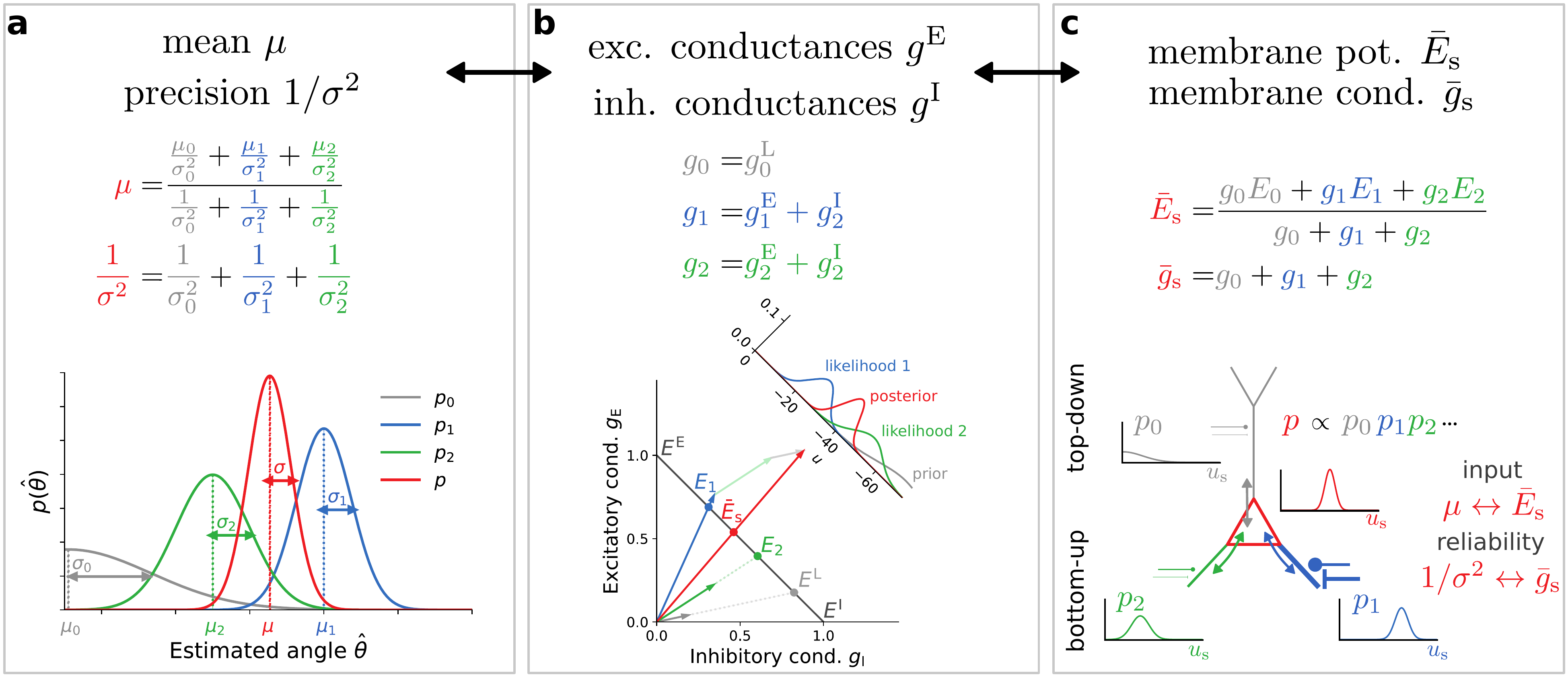}
    \caption{{\bf Non-linear \replaced[id=JJ]{cue integration}{opinion weighting} is achieved through a linear vector summation of conductances.}
      {\bf (a)} Non-linear combination of Gaussian probability densities.
      The pooled mean is a convex combination of the original means, while the pooled reliability, the inverse variance, is a sum of the individual reliabilities.
      {\bf (b)} Stimulus-evoked excitatory and inhibitory synaptic conductances as two-dimensional vectors (blue and green), as well as the leak (gray), are linearly summed across dendrites to yield the total somatic conductances (red arrow).
      The intersections with the antidiagonal (black line) yield the corresponding dendritic and somatic reversal potentials.
      This intersection is a nonlinear operation (see Methods Sec. "\nameref{sec:methods-linear-coordinates-for-nonlinear-processing}").
      The inset shows the full distributions.
      Note that the prior can be modulated by synaptic conductance elicited by top-down input (see panel c).
      {\bf (c)} Translation of prior (gray) and dendritic (green and blue) \replaced[id=JJ]{potentials and conductances}{opinions and reliabilities} into the corresponding somatic mean potential and conductances (red).
      For visualization purposes, the prior distribution is only partially shown.
    }\label{fig:results-math-to-neurons}
\end{figure}

\subsection*{Bayesian neuronal dynamics}
\added[id=JJ]{Excitatory and inhibitory conductances targeting a neuronal compartment combine with the leak and the associated reversal potentials into a total transmembrane current $I^\text{d} = \gd \, \left( \Ed - \ud \right)$.
This current induces a stimulus-dependent effective reversal potential $\Ed$ given by}
\begin{align}
\Ed = \frac{\gE \EE + \gI \EI + \gL \EL}{\gE + \gI + \gL} \; ,
\label{eq:results-Ed}
\end{align}
\added[id=JJ]{where excitatory, inhibitory and leak reversal potential are denoted as $E^\text{E/I/L}$, and the respective conductances by $g^{E/I/L}$.
The sum of these three conductances $\gd = \gE + \gI + \gL$ represents the local membrane conductance, which excludes the coupling to other compartments.
The excitatory and inhibitory conductances are the product of the synaptic weights times the presynaptic firing rates, $g^\text{E/I} = W^\text{E/I} r$.
Note that in general $\Ed$ is different from the actual dendritic potential $\ud$, which is additionally influenced by the membrane potential in neighboring compartments.}

\added[id=JJ]{Across the dendritic tree we now interpret $\gd_i$ and $\Ed_i$ as parameters of Gaussian \cite{Petersen2016} likelihood functions $p(\Ed_i | \us, \gd_i)$ in basal compartments and parameters of Gaussian priors $p(\us | \Ed_i, \gd_i)$ in apical compartments.
The dendritic likelihoods quantify the statistical relationship between dendritic and somatic potentials.
Intuitively speaking, they describe how compatible a certain somatic potential $\us$ is with an effective reversal potential $\Ed_i$.
Note that this relation is of purely statistical, not causal nature -- biophysically, effective reversal potentials $\Ed_i$ cause somatic potentials, not the other way around.}

Finally, the somatic compartment computes the posterior according to Bayes theorem (see Methods Sec.~"\nameref{sec:methods-bayesian-theory-of-somatic-potential-dynamics}" for details),
\begin{align}
    p(\us | W, r) \propto \text{likelihood} \times \text{prior} = e^{-\frac{\bargs}{2 \lambdae}(\us - \barEs)^2} \; .
    \label{eq:results-probability-density-soma}
\end{align}
\added[id=JJ]{Here, $\bargs$ represents the total somatic conductance, and $\barEs$ the total somatic reversal potential, which is given by the convex combination of the somatic and dendritic effective reversal potentials, weighted by their respective membrane conductances and dendro-somatic coupling factors (Fig.~\ref{fig:results-math-to-neurons}).
The "exploration parameter" $\lambdae$ relates conductances to membrane potential fluctuations.
In general, this parameter depends on neuronal properties, for example, on the amplitude of background inputs and the spatial structure of the cell.
It can be determined experimentally by an appropriate measurement of membrane potentials from which both fluctuation amplitudes and decay time constants $\tau = C/\bargs$ can be estimated.}

\added[id=JJ]{To obtain the somatic membrane potential dynamics, we propose that the soma performs noisy gradient ascent on the log-posterior,}
\begin{align}
  C \dotus = & \; \lambdae \frac{\partial}{\partial \us} \log p(\us| W,r) + \xi \notag \\
           = & \bargs \, (\barEs - \us ) + \xi \notag \\
           = & \gprior (\Eprior - \us) + \sum_{i=1}^D \alphasd_i \left[ g_i^\Lk (\EL - \us) + g_i^\E (\EE - \us) + g_i^\I (\EI - \us) \right] + \xi \; .
  \label{eq:results-somatic-dynamics}
\end{align}
\added[id=JJ]{with membrane capacitance $C$, and dendro-somatic coupling factors $\alphasd_i = \gsdc_i / (\gsdc_i + \gd_i)$ that result from the dendro-somatic coupling conductances $\gsdc_i$ and the isolated dendritic conductances $\gd_i$.}
\added[id=JJ]{The additive noise $\xi$ represents white noise with variance $2 C \lambdae$, arising, for example, from unspecific background inputs \cite{richardson2005synaptic,petrovici2016stochastic,dold2019stochasticity,jordan2019deterministic}.
For fixed presynaptic activity $r$, the average somatic membrane potential hence represents a maximum-a-posteriori estimate (MAP, \cite{knill2004bayesian}), while its variance is inversely proportional to the total somatic conductance $\bargs$.
The effective time constant of the somatic dynamics is $\tau = C / \bargs$, thus enabling $\us$ to converge faster to reliable MAP estimates for larger $\bargs$.}

\added[id=JJ]{The dynamics derived here from Bayesian inference  (Eqn.~\ref{eq:results-somatic-dynamics}) are identical to the somatic membrane potential dynamics in bidirectionally coupled multi-compartment models with leaky integrator dynamics and conductance-based synaptic coupling under the assumption of fast dendritic responses\cite{wybo2019electrical}.
In other words, the biophysical system computes the posterior distribution via its natural evolution over time.
This suggests a fundamental role of conductance-based dynamics for Bayesian neuronal computation.}
%
%

\added[id=JJ]{Conductance-based Bayesian integration, as introduced above, can also be viewed from a different perspective in terms of probabilistic opinion pooling \cite{dietrich2016probabilistic}.
Under this view each dendrite can be thought of as an individual with a specific opinion -- the dendrite's effective reversal potential -- along with an associated reliability -- the dendrite's conductance.
Accordingly, the soma then plays the role of a "decision maker" that pools the reliability-weighted dendrite's opinions, determines a compromise, and communicates this outcome to other individuals, i.e., downstream neurons' dendrites.
Intuitively speaking, in this process dendrites with a lot of confidence in their opinion, i.e., those with high dendritic conductance, contribute more to the pooled opinion than others.
}

\added[id=JJ]{Before introducing synaptic plasticity, we first discuss a specific consequence for neuronal dynamics arising from our Bayesian view of neuronal dynamics.}

\subsection*{\deleted[id=JJ]{The neuronal opinion code}}
\deleted[id=JJ]{Excitatory and inhibitory conductances targeting a dendritic compartment combine with the dendritic leak and the associated reversal potentials into a total dendritic transmembrane current $I^\text{d} = \gd \, \left( \Ed - \ud \right)$.
Here, the local, stimulus-dependent dendritic reversal potential $\Ed$ is given by}
\deleted[id=JJ]{where excitatory, inhibitory and leak reversal potential are denoted as $E^\text{E/I/L}$, and the respective conductances by $g^{E/I/L}$.
The sum of these three conductances $\gd = \gE + \gI + \gL$ represents the isolated dendritic conductance, which excludes the somato-dendritic coupling.
The excitatory and inhibitory conductances are the product of the synaptic weights times the presynaptic firing rates, $g^\text{E/I} = W^\text{E/I} r$.
Note that in general $\Ed$ is different from the actual dendritic potential $\ud$, which is additionally influenced by the somatic potential.}

\deleted[id=JJ]{In our framework, each dendritic compartment has an associated preferred feature, i.e., an activity pattern in its afferents which maximizes its reversal potential $\Ed$.
We hence identify $\Ed$ with the dendritic opinion about how well presynaptic activity is compatible with its preferred feature.
We furthermore identify the isolated dendritic conductance $\gd$ with the reliability of the corresponding dendritic opinion.
Intuitively speaking, the opinion of a dendritic compartment with large dendritic conductance will be more resilient against different opinions encoded in other compartments.}

\deleted[id=JJ]{How are dendritic opinions pooled to jointly determine the output of a neuron?
The interaction between soma and dendrites in cortical neurons naturally form a pooled opinion $\barEs$ as a weighted average of the individual dendritic opinions $\Ed_i$, with the weight of each dendritic opinion reflecting its reliability $\gd_i$ (Fig.~\ref{fig:results-math-to-neurons}c).
The reliability of this pooled opinion is reflected by the total somatic conductance $\bargs$.
The somatic membrane potential $\us$ dynamically traces a noisy estimate of the pooled opinion $\barEs$,}
\deleted[id=JJ]{with membrane capacitance $C$, and dendro-somatic coupling factors $\alphasd_i = \gsdc_i / (\gsdc_i + \gd_i)$ that result from the dendro-somatic coupling conductances $\gsdc_i$ and the isolated dendritic conductances $\gd_i$ (see Methods, "Bayesian theory of somatic potential dynamics" for details).}

\deleted[id=JJ]{The corresponding neuronal processing of inputs is a non-linear operation on the level of membrane potentials, described by sublinear summations \cite{Tran-Van-Minh2015}.
Despite the nonlinear effect of inputs on membrane potentials, the operations in conductance space are described by purely linear operations.
While the dendritic sublinearity gives rise to dendritic opinions, the somatic sublinearity performs a Bayesian combination of opinions across the dendritic tree (Fig.~\ref{fig:results-math-to-neurons}a,c).}

\deleted[id=JJ]{In our framework the somatic membrane potential is transformed into an instantaneous output rate $\rs = \rho(\us)$ via some monotonically increasing transfer function $\rho$.
At any given time, multiple neurons with identical preferred features will produce different output rates due to random background input.
The variance across such an ensemble reflects the reliability of the somatic opinions and can thus be communicated to downstream areas (Fig.~\ref{fig:results-cond-neuron-sketch}f) which can adjust synaptic weights to take this variability into account.}

\subsection*{\replaced[id=JJ]{Stimuli lead to Bayesian updates of somatic membrane potential statistics}{Stimuli elicit neuron-specific opinions and increase the neuronal reliability}}
\begin{figure}[tbp!]
    \centering
    \includegraphics[width=1.\textwidth]{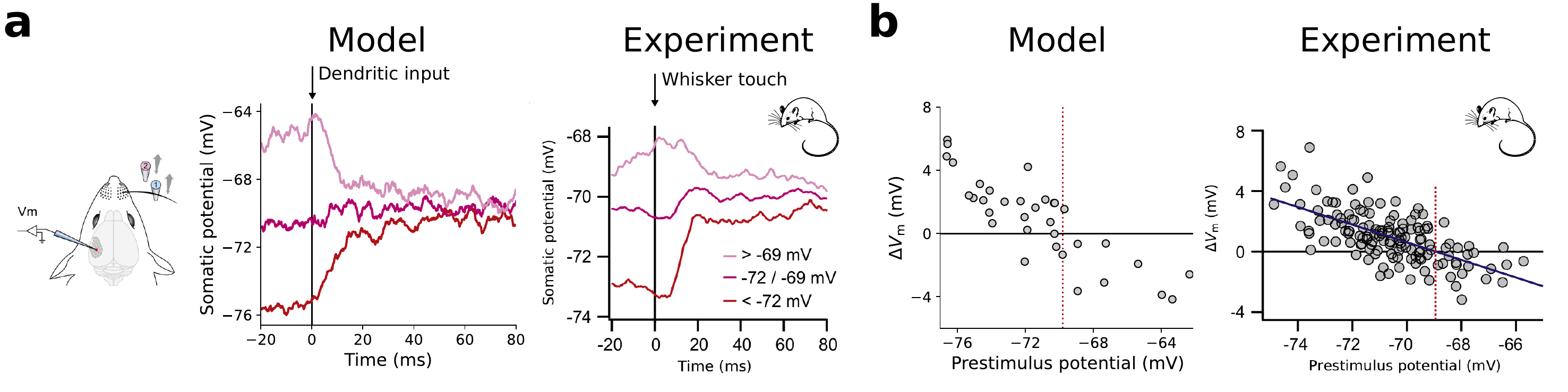}
    \caption{
    {\bf \replaced[id=JJ]{Conductance-based Bayesian integration}{Dendritic opinion pooling} implies stimulus-specific reversal potentials.}
    {\bf (a)} Average stimulus-evoked responses for different ranges of prestimulus potentials generated by our model (left) and measured experimentally (right, from \cite{crochet2011synaptic}).
    Vertical arrow indicates stimulus onset corresponding to activation of dendritic input and whisker touch, respectively.
    Independently of the previous value of the somatic potential, the dendritic input always pulls the somatic potential towards the effective reversal potential associated with the stimulus.
    {\bf (b)} PSP amplitude vs.~prestimulus potential generated by our model (left) and measured experimentally (right, from \cite{crochet2011synaptic}).
    Reprinted from Neuron, 69, Crochet, S., Poulet, J. F., Kremer, Y. \& Petersen, C. C., Synaptic mechanisms underlying sparse coding of active touch, 1160--1175, Copyright (2011), with permission from Elsevier.
  }\label{fig:results-reversal-potentials0}
\end{figure}

The conductance-\replaced[id=JJ]{based Bayesian integration view}{centered neuronal opinion weighting framework} predicts neuronal response properties that differ from those of classical \deleted[id=JJ]{current-based }neuron models.
\replaced[id=JJ]{In the case of conductances, somatic membrane potentials reflect prior expectations}{In the opinion weighting framework, prior opinions are encoded in the somatic membrane potential} in the absence of sensory input.
These priors typically have low reliability, encoded in relatively small conductances.
As a consequence, the neuron is more susceptible to background noise, resulting in large membrane potential fluctuations\deleted[id=JJ]{ around the prior potential}.
\replaced[id=JJ]{Upon stimulus onset}{When a cue is presented}, presynaptic activity increases causing synaptic conductances to increase, thereby pulling postsynaptic membrane potentials towards the cue-specific reversal potentials $\Ed$, irrespective of their prior value (Fig.~\ref{fig:results-reversal-potentials0}a).
This phenomenon is observed in electrophysiological recordings from mouse somatosensory cortex: the change in membrane potential upon whisker stimulation pulls the somatic membrane potential from variable \deleted[id=JJ]{spontaneous }pre-stimulus potentials\added[id=JJ]{, i.e., different prior expectations,} towards a cue-specific post-stimulus potential (Fig.~\ref{fig:results-reversal-potentials0}a, \cite{crochet2011synaptic}).
Besides a change in the average membrane potential, cue onset increases conductances and hence decreases \replaced[id=JJ]{membrane potential variability}{spontaneous fluctuations}.

These effects are signatures of \replaced[id=JJ]{Bayesian computations}{neuronal opinion weighting}.
\deleted[id=JJ]{Cues provide information about the presence or absence of a neuron's preferred feature.}
Upon cue onset, the prior distribution \deleted[id=JJ]{(i.e., distribution in the absence of cues) }is combined with \replaced[id=JJ]{stimulus-specific likelihoods}{cue-specific distributions} leading to an updated somatic distribution with adapted mean and reduced variance.
If the prior strongly disagrees with \replaced[id=JJ]{information provided by the stimulus}{cue information}, the change in mean is larger than if prior and \replaced[id=JJ]{stimulus}{cue} information are consistent.
Importantly, the variance is always reduced in the presence of new information, regardless of whether it conflicts with previous information or not; this is a hallmark of Bayesian reasoning.

We propose that this probabilistic computation underlies the observed stimulus-driven reduction of variability throughout cortex \cite{monier2003orientation,churchland2010stimulus} and explains why stimulus-evoked PSP amplitudes are negatively correlated with prestimulus potentials \cite[Fig.~\ref{fig:results-reversal-potentials0}b; also see][]{crochet2011synaptic,sachidhanandam2013membrane}.
In whisker stimulation experiments \cite{crochet2011synaptic}, the stimulation intensity is encoded by the whisker deflection angle.
Our framework predicts that, as the amplitude of whisker deflections increases, the variance of the post-stimulus potentials decreases.
This prediction is consistent with the recent observation that increasing the contrast of oriented bar stimuli reduces the variance in the postsynaptic response of orientation-specific neurons in macaque visual cortex \cite{henaff2020representation}.
Furthermore, our model predicts that the nature of stimuli during learning will affect the impact of sensory cues on electrophysiological quantities and behavior: more reliable priors will cause a smaller influence of sensory inputs, while increasing stimulus reliability\added[id=JJ]{, e.g., stimulus intensity,} would achieve the opposite effect.
Regardless of training, our model also predicts decreasing influence of the prior for increasing stimulus intensity.

\subsection*{\deleted[id=JJ]{Bayesian neuronal dynamics}}

\deleted[id=JJ]{The proposed neuronal opinion weighting can be described in a probabilistic framework of neuronal coding.
This framework allows us to derive the same biophysical dynamics, but from a normative standpoint.}

\deleted[id=JJ]{For given synaptic weights $W$ and presynaptic rates $r$ that encode information about sensory stimuli, we propose that the soma computes a posterior distribution over its membrane potential $p(\us | W, r)$.
Absent any sensory input, we assume the somatic potential follows a Gaussian prior $p(\us | \Eprior, \gprior)$.
Its mean represents the prior somatic opinion $\Eprior$ and its variance is the inverse of the prior reliability $\gprior$ (cf. Fig.~\ref{fig:results-math-to-neurons}); these parameters are determined by a combination of leak and non-sensory (top-down or lateral) inputs.
Consistent with experimental data \cite{Petersen2016} we assume Gaussian dendritic likelihoods $p(\Ed_i | \us, \gd_i)$ with dendritic reversal potentials $\Ed_i$ and isolated dendritic conductances $\gd_i$ determined by synaptic weights and presynaptic rates as discussed above.
The dendritic likelihoods quantify the statistical relationship between dendritic opinions and the somatic potential.
Intuitively speaking, they describe how compatible a certain somatic potential $\us$ is with a dendritic reversal potential $\Ed_i$.
Note that this relation is of purely statistical, not causal nature -- biophysically, dendritic reversal potentials $\Ed_i$ cause somatic potentials, not the other way around.
To perform probabilistic inference, the soma computes the posterior via Bayes' theorem:}
\deleted[id=JJ]{Here, $\bargs$ represents the total somatic conductance, and $\barEs$ the pooled somatic opinion, which is given by the convex combination of the somatic and dendritic opinions, weighted by their respective reliabilities and dendro-somatic coupling factors (see Methods, "Bayesian theory of somatic potential dynamics" and Fig.~\ref{fig:results-math-to-neurons}).
The exploration parameter $\lambdae$ relates conductances to membrane potential fluctuations.
In general, this parameter depends on neuronal properties, for example, on the amplitude of background inputs and the spatial structure of the cell.}

\deleted[id=JJ]{To obtain the somatic membrane potential dynamics from its statistics, we postulate that the soma performs noisy gradient ascent on the log-posterior of the somatic potential:}
\deleted[id=JJ]{The additive noise $\xi$ represents white noise with variance $2 C \lambdae$, arising, for example, from unspecific background inputs \cite{richardson2005synaptic,jordan2019deterministic}.
For fixed presynaptic activity $r$, the average somatic membrane potential hence represents a maximum-a-posteriori estimate (MAP, \cite{knill2004bayesian}), while its variance is inversely proportional to the total somatic conductance $\bargs$.
The effective time constant of the somatic dynamics is $\tau = C / \bargs$, thus enabling $\us$ to converge faster to reliable MAP estimates for larger $\bargs$.}

\deleted[id=JJ]{The dynamics derived here from Bayesian inference are identical to the somatic membrane potential dynamics in bidirectionally-coupled multi-compartment models with leaky integrator dynamics and conductance-based synaptic coupling under the assumption of fast dendritic responses (Eqn.~\ref{eq:results-somatic-dynamics}).
In other words, the biophysical system effectively computes the posterior distribution via its natural evolution over time.
This suggests a fundamental role of conductance-based dynamics for Bayesian neuronal computation, which also extends to synaptic plasticity, as we discuss in the following.}

\begin{figure}[tbp]
    \centering
    \includegraphics[width=.75\textwidth]{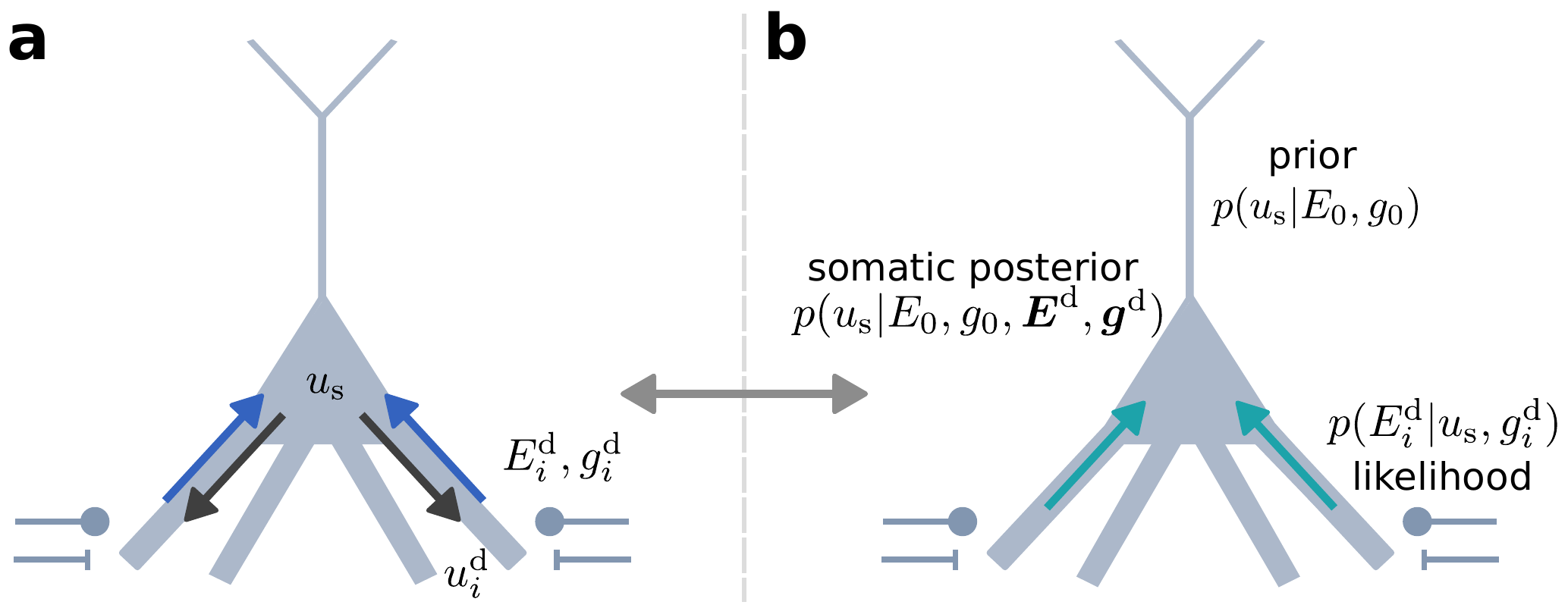}
    \caption{{\bf Single neuron dynamics as Bayesian inference.}
    {\bf (a)} \replaced[id=JJ]{Somatic and dendritic membrane potentials are coupled through currents flowing along the dendritic tree}{Biophysical dynamics bidirectionally couple somatic and dendritic membrane potentials} (\added[id=JJ]{blue and black arrows, }Eqs~\ref{Seq:duv}, \ref{eq:Sduv}).
    {\bf (b)} The\deleted[id=JJ]{ir} steady state \added[id=JJ]{of the somatic compartment }can be interpreted as computing the posterior $p(\us \,|\, \Eprior, \gprior, \vecEd, \vecgd)$ from the \replaced[id=JJ]{dendritic}{somatic} priors $p( u_\text{s} | \Eprior, \gprior)$ and dendritic likelihoods $p(\Ed_i|\us,\gd_i)$.
    \replaced[id=JJ]{Stimulus-driven effective reversal potentials in basal dendrires pull}{The current flow between dendrites and soma pulls} the somatic potential distribution from the prior towards the posterior\deleted[id=JJ]{, reflecting the update of the somatic opinion by the properly weighted dendritic opinions}.
  }\label{fig:SI_Fig1}
\end{figure}

\subsection*{Gradient-based synaptic dynamics}

\replaced[id=JJ]{As discussed above, a}{A} fixed stimulus determines the somatic membrane potential distribution\deleted[id=JJ]{ and the somatic membrane potential dynamics will continuously sample from this distribution}.
Prior to learning, this distribution will typically be different from a desired distribution as predicted, for example, by past sensory experience or cross-modal input.
We refer to such \replaced[id=JJ]{stimulus}{input}-dependent desired distributions as target distributions.

We define learning in our framework as adapting synaptic weights $W$ to \replaced[id=JJ]{increase the probability of samples $\us^*$ from the target distribution under the currently represented somatic posterior}{increase the probability of observing samples $\us^*$ from the target distribution}.
Formally, learning reduces the Kullback-Leibler divergence \added[id=JJ]{$KL(p^*|p)$} between the target distribution $p^*(\us|r)$ and the somatic \added[id=JJ]{membrane potential} distribution $p(\us | W, r)$.
\added[id=JJ]{This can be interpreted as a form of supervised learning, where a large divergence implies poor performance and a small divergence good performance, respectively.}
This is achieved through gradient ascent on the (log-)posterior somatic probability of target potentials $\us^*$ sampled from the target distribution, resulting in the following dynamics for excitatory and inhibitory weights (for details see Methods Sec.~"\nameref{sec:methods-weight-dynamics}"):
\begin{align}
\dot W_i^\text{E/I} \propto \lambdae \frac{\partial}{\partial W_i^\text{E}} \log p(\us^*|W,r)
  & \propto \bigg[ \underbrace{(\us^* - \barEs) \left( E^\text{E/I} - \tildeEd_i \right)}_{=\Delta \mu^\text{E/I}_i} + \underbrace{\frac{\alphasd_i}{2} \left( \frac{\lambdae}{\bargs} - (\us^* - \barEs)^2 \right)}_{=\Delta \sigma^2} \bigg] \, r \;, 
     \label{eq:dWd0}
\end{align}
with $\tildeEd_i = \alphasd_i \barEs + (1-\alphasd_i) \Ed_i$. Here, $\lambdae$ is the exploration parameter, $\alphasd_i$ the an effective dendritic coupling strength, $\Ed_i$ the reversal potential of dendrite $i$ given by Eqn.~\ref{eq:results-Ed}, and $\barEs$ the \replaced[id=JJ]{total somatic reversal potential}{mean somatic potential that can be seen as an effective reversal potential (see Eq.)}.

All dynamic quantities arising in the synaptic plasticity rule are neuron-local.
The dendritic potentials $\Ed_i$ are available at the synaptic site, as well as the presynaptic rates $r$.
We hypothesize that the backpropagating action potential rate that codes for $\us^*$ can influence dendritic synapses \cite{Urbanczik2014}.
Furthermore, the total conductance $\bargs$ determines the effective time constant by which the somatic membrane potential fluctuates and could be measured through its temporal correlation length.
The exact molecular mechanisms by which these terms and their combinations are computed in the synapses remain a topic for future research.

\subsection*{Joint learning of somatic mean and variance}

\begin{figure}[tbp]
  \centering
  \includegraphics[width=0.75\textwidth]{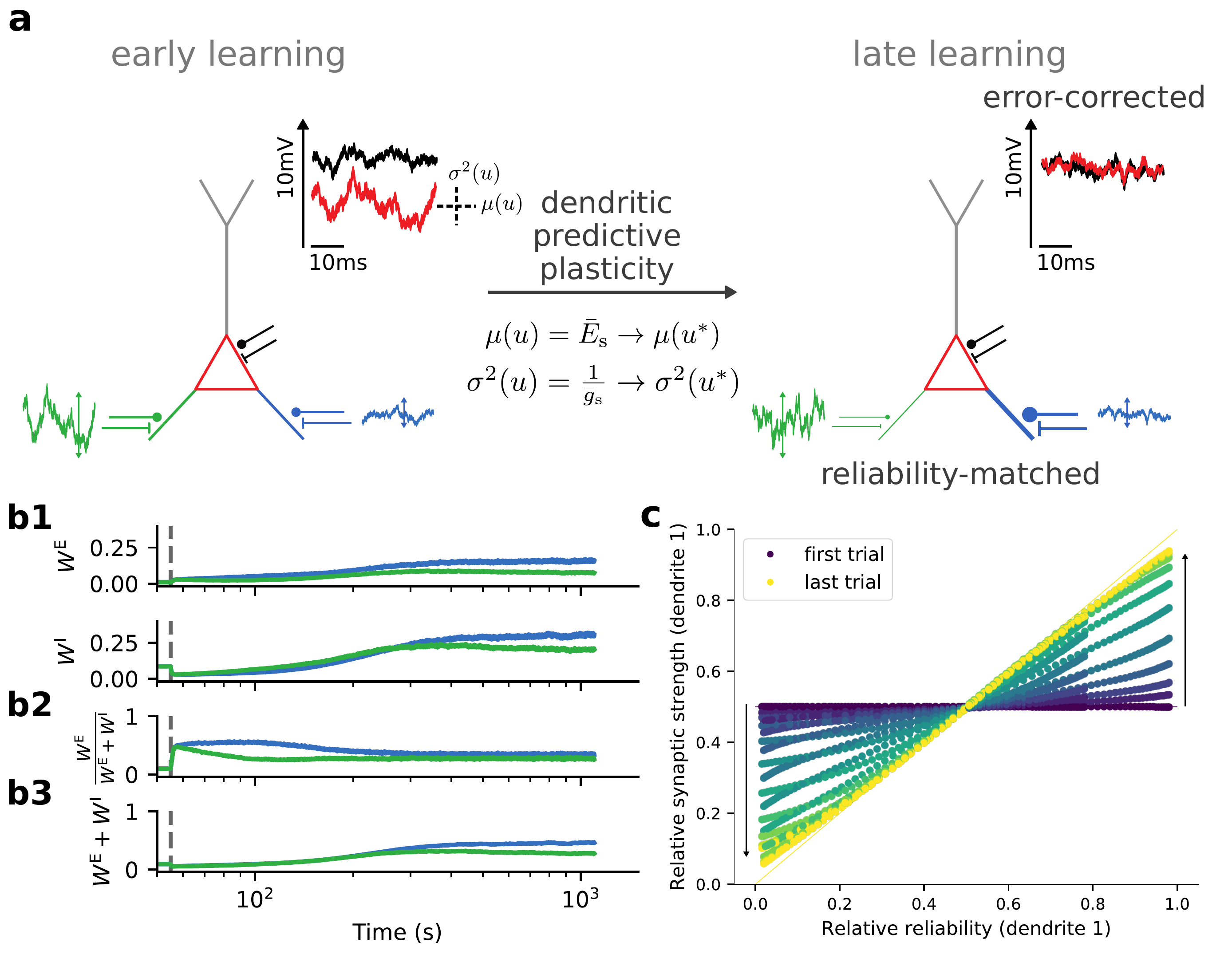}\\
  \caption{{\bf Dendritic predictive plasticity performs error correction and reliability matching.}
    {\bf (a)} A neuron receives input via two different input channels with different noise amplitudes (green and blue).
    Synaptic plasticity adapts the mean ($\mu$) and variance ($\sigma^2$) of the somatic membrane potential (red) towards the target (black).
    {\bf (b1)} Excitatory and inhibitory weights per input channel (basal dendrite).
    The dashed vertical line indicates the onset of learning.
    The dendrites learn the mean target potential within the first few seconds (jumps after the dashed line).
    {\bf (b2)} Ratio of excitatory and total synaptic weights per dendrite. These ratios determine the mean dendritic membrane potentials. Since both dendrites learn to match the same somatic mean potential based on their respective synaptic inputs, these ratios become equal.
    {\bf (b3)} Sum of excitatory and inhibitory weights per dendrite. The total dendritic weights reflect the reliability of the dendritic \replaced[id=JJ]{input}{opinion}.
    Learning assigns larger synaptic weights to the less fluctuating and more reliable input (blue) as compared to the stronger fluctuating and less reliable input (green).
    As the balancing ratio becomes the same (b2), the excitatory and inhibitory strengths of the more reliable input must both become larger (b1).
    {\bf (c)} The relative synaptic strength of a given branch ($W_i/\sum_j W_j$) becomes identical to the relative reliability ($\frac{1}{\sigma_i^2}/\sum_j \frac{1}{\sigma_j^2}$) of its input with respect to the other branches over the course of learning (here shown for $i=1$; starting with $W_1=W_2$ for the entire range of relative reliabilities, horizontal line).
    Note that time flows from blue (first trial) to yellow (last trial).
  }
  \label{fig:results-plasticity}
\end{figure}

The total postsynaptic error is composed of an error in the mean $\Delta \mu^\text{E/I}_i$ and an error in the variance $\Delta \sigma^2$ (Eqn.~\ref{eq:dWd0}).
By jointly adapting the excitatory and inhibitory synapses, both errors in the mean and the variance are reduced.
To simultaneously adjust both the mean and variance, the two degrees of freedom offered by separate excitation and inhibition are required.

To illustrate these learning principles we consider a toy example in which a neuron receives input via two different input channels with different noise amplitudes.
Initially neither the average somatic membrane potential, nor its variance match the \added[id=JJ]{the parameters of the }target distribution (Fig.~\ref{fig:results-plasticity}a, left).
Over the course of learning, the ratio of excitatory to inhibitory weights increases to allow the \added[id=JJ]{average }somatic membrane potential to match the average target potential and the total strength of both excitatory and inhibitory inputs increases to match the \added[id=JJ]{inverse of the total somatic conductance to the }variance of the targets (Fig.~\ref{fig:results-plasticity}a, right; b$1$).
Excitatory and inhibitory weights hence first move into opposite directions to match the average, and later move in identical directions to match the variance (Fig.~\ref{fig:results-plasticity}b$1$).

In both dendrites, the strengths of excitation and inhibition converge to the same ratio to match the mean of the target distribution (Fig.~\ref{fig:results-plasticity}b$2$).
However, the relative magnitude of the total synaptic strength $W^\text{tot}  = W^\text{E} + W^\text{I}$ changes according to the relative fluctuations of the presynaptic input during learning.
While branches with reliable presynaptic input (small fluctuations) are assigned large total synaptic weights, branches with unreliable input learn small total synaptic weights (Fig.~\ref{fig:results-plasticity}b2).
More specifically, the total synaptic weights indeed match the respective reliabilities of the individual dendrites: $W^\text{tot} \appropto \frac{1}{\sigma_r^2}$ (Fig.~\ref{fig:results-plasticity}c).
\replaced[id=JJ]{Intuitively}{Roughly} speaking, the total synaptic weights learn to modulate \replaced[id=JJ]{somatic background}{the somatic} noise $\xi$ towards a target variance $\sigma_u^*$.
\deleted[id=JJ]{However, since the inputs themselves are unreliable, their influence needs to be modulated by this unreliability, i.e., by the variance $\sigma_r^2$.}
For a proof, we refer to the SI.


\subsection*{Learning Bayes-optimal cue combinations}

\begin{figure}[tbp]
    \centering
    \includegraphics[width=1.\textwidth]{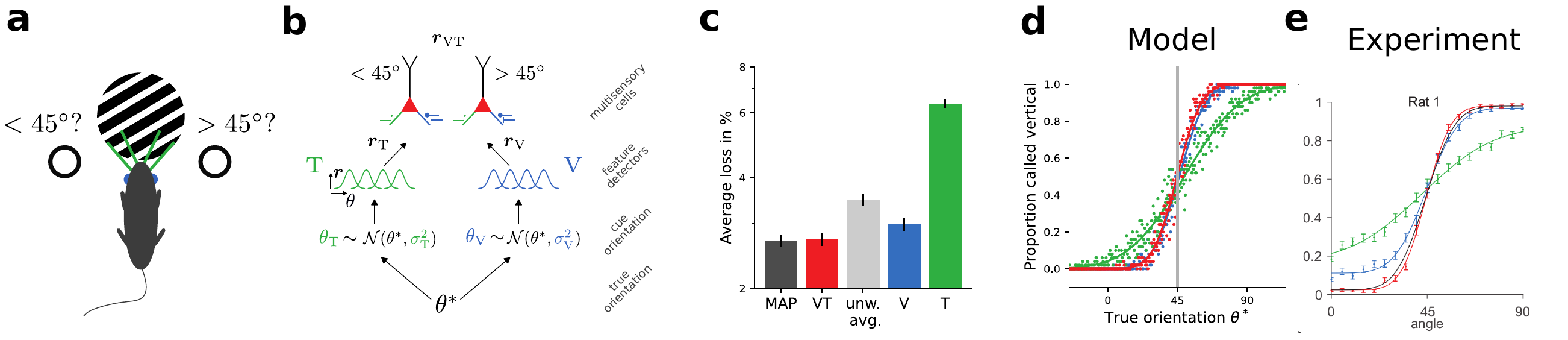}
    \caption{{\bf Learning Bayes-optimal inference of orientations from multimodal stimuli.}
        {\bf (a)} Experimental setup \cite[see also][]{nikbakht2018supralinear}.
        {\bf (b)} Network model.
        {\bf (c)} Accuracy of the MAP estimate (MAP, dark gray), the trained model with bimodal cues (VT, red), unweighted average of visual and tactile cues (unw.~avg., light gray), and the trained model with only visual (V, blue) and tactile cues (T, green), respectively.
        Error bars denotes standard error of the mean over $25$ experiments, each consisting of $20\,000$ trials.
        The trained model performs as well as a theoretically optimal observer (compare loss of MAP and VT).
        {\bf (d)} Psychometric curves of the model confirm that the classification near $45^\circ$ for the combined modalities (red) is at least as good as for the visual modality (V, blue, lower input variance), and better than for the tactile modality (T, green, higher input variability).
        Dots: subsampled data, solid lines: fit of complementary error function.
        {\bf (e)} Psychometric curves for rat $1$ \cite{nikbakht2018supralinear} for comparison.
        Reprinted from Neuron, 97, Nikbakht, N., Tafreshiha, A., Zoccolan, D. \& Diamond, M. E., Supralinear and supramodal integration of visual and tactile signals in rats: psychophysics and neuronal mechanisms, 626--639, Copyright (2018), with permission from Elsevier.
    }\label{fig:results-orientation-estimation}
\end{figure}

We next consider a multisensory integration task in which a rat has to judge whether the angle of a grating is larger than $45^\circ$ or not, using whisker touching (T) and visual inspection (V), see Fig.~\ref{fig:results-orientation-estimation}a and \cite{nikbakht2018supralinear}.
In this example, projections are clustered according to modality on dendritic compartments.
In general, this clustering is not necessarily determined by modality but could also reflect, for example, lower-level features, or specific intracortical pathways.
In our setup, uncertainty in the sensory input from the two modalities is modeled by different levels of additive noise.
The binary classification is performed by two multisensory output neurons that are trained to encode the features $>45^\circ$ and $<45^\circ$, respectively.
Technically, we assume the target distribution is a narrow Gaussian centered around a stimulus-dependent target potential.
For example, for the neuron encoding orientations $>45^\circ$, the target potential would be high for ground truth orientations $>45^\circ$ and it would be low otherwise.
The output neurons receive input from populations of feature detectors encoding information about visual and tactile cues, respectively (Fig.~\ref{fig:results-orientation-estimation}b).

The performance of the model neurons after learning matches well the Bayes-optimal MAP estimates that make use of knowledge about the exact relative noise variances.
In contrast, averaging the two cues with equal weighting, and thus not exploiting the conductance-based \replaced[id=JJ]{Bayesian processing}{opinion pooling}, or considering only one of the two cues, would result in lower performance (Fig.~\ref{fig:results-orientation-estimation}c).
Furthermore, the psychophysical curves of the trained model match well to experimental data obtained in a comparable setup (Fig.~\ref{fig:results-orientation-estimation}d,e).

\subsection*{Cross-modal suppression is caused by \replaced[id=JJ]{conductance-based Bayesian integration}{reliability-weighted opinions}}

\begin{figure}[tbp]
    \centering
    \includegraphics[width=1.\textwidth]{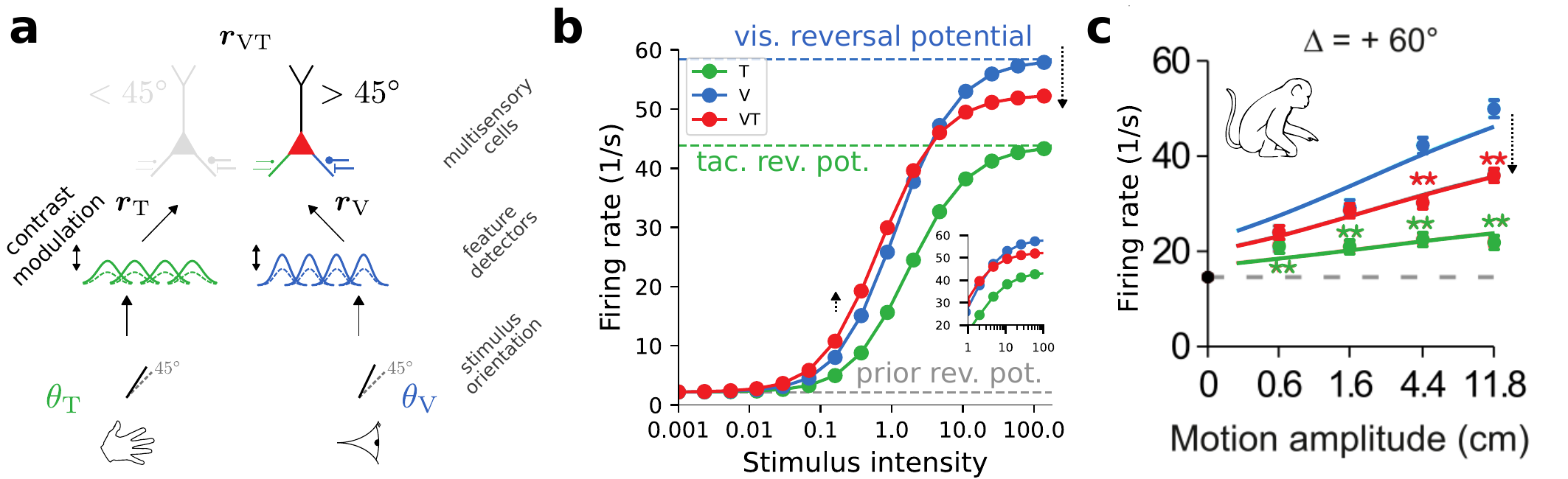}
    \caption{{\bf Cross-modal suppression \replaced[id=JJ]{arising from Bayes-optimal integration of information in single neurons}{as reliability-weighted opinion pooling}.}
    {\bf (a)} Experimental setup (compare Fig.~\ref{fig:results-orientation-estimation}).
    {\bf (b)} Firing rate of the output neuron encoding orientations $>45^\circ$ for unimodal stimulation (V,T) and bimodal stimulation (VT).
    Dashed lines indicate the limit of no stimulation (gray), and infinitely strong tactile (green) and visual (blue) stimulation, respectively.
    Inset shows zoom in for high stimulation intensities.
    Pulling the somatic \replaced[id=JJ]{potential}{opinion} (red) towards the weighted mean of the visual and tactile \replaced[id=JJ]{effective reversal potentials}{opinion} (blue and green dashed lines) leads to a relative increase for weak stimulus intensities (black upward arrow) and to cross-modal suppression at strong stimulus intensities (black downward arrow).
    {\bf (c)} Firing rate of a neuron from macaque MSTd in response to misaligned visual (blue) and vestibular (green) cues with a mismatch of $\Delta = 60^\circ$.
    Modified from~\cite{ohshiro2017neural}.
    Reprinted from Neuron, 95, Ohshiro, T., Angelaki, D. E. \& DeAngelis, G. C., A neural signature of divisive normalization at the level of multisensory integration in primate cortex, 399--411, Copyright (2017), with permission from Elsevier.
  }\label{fig:results-crossmodal-suppression}
\end{figure}

Using the trained network from the previous section, we next consider the firing rate of the output neuron that prefers orientations $> 45^\circ$ for conflicting cues with a specific mismatch.
We assume a true stimulus orientation $> 45^\circ$ generates a separate cue for each modality, where, as an example we assume the visual cue to be more vertical than the tactile cue (Fig.~\ref{fig:results-crossmodal-suppression}a) which result in different dendritic reversal potentials $\Ed_i$.
In the following we identify the reliability of a stimulus with its intensity.
Intuitively speaking, a weak stimulus is less reliable than a strong one.

When cues are presented simultaneously at low stimulus intensity, the output neurons fire stronger than in unimodal conditions (Fig.~\ref{fig:results-crossmodal-suppression}b).
However, when presented simultaneously at high stimulus intensity the cues suppress each other, i.e., the resulting firing rate is smaller than the maximal rate in unimodal conditions (Fig.~\ref{fig:results-crossmodal-suppression}b).
This phenomenon is known as cross-modal suppression \cite{fetsch2013bridging,ohshiro2017neural}.

In the context of the \replaced[id=JJ]{conductance-based Bayesian integration}{opinion weighting}, this counterintuitive interaction of multimodal cues arises as a consequence of the \replaced[id=JJ]{somatic potential}{pooled opinion} being a weighted average of the two unimodal \replaced[id=JJ]{effective reversal potentials}{opinions} and the prior\deleted[id=JJ]{ opinion}.
For low stimulus intensity the prior \deleted[id=JJ]{opinion }dominates; since the evidence from either modality is only weak, \replaced[id=JJ]{information}{the opinion} arriving from a second modality always constitutes additional evidence that the preferred stimulus is present.
Thus, the \replaced[id=JJ]{somatic potential}{pooled opinion} is pulled farther away from the prior in the bimodal condition as compared to the unimodal one.
For high stimulus intensity the prior does not play a role and the \replaced[id=JJ]{somatic potential}{pooled opinion} becomes a weighted average of the two modality-specific \replaced[id=JJ]{effective reversal potentials}{opinions}.
As one cue is more aligned with the neuron's preferred feature than the other, the weighted average appears as a suppression (Fig.~\ref{fig:results-crossmodal-suppression}).

We \deleted[id=JJ]{thus }propose that the computational principle of \replaced[id=JJ]{conductance-based Bayesian integration also}{dendritic opinion pooling} underlies other \replaced[id=JJ]{variants}{versions} of cross-modal suppression \cite[e.g.,][]{carandini1994summation,fetsch2013bridging,ohshiro2017neural,meijer2017audiovisual}, and also explains unimodal suppression arising from superimposing cues \cite[e.g.,][]{morrone1982functional,carandini1997linearity,busse2009representation}, or superimposing sensory inputs and optogenetic stimulation \cite{sato2014distal,nassi2015optogenetic}.

\section*{Discussion}

The biophysics of cortical neurons can be interpreted \replaced[id=JJ]{as Bayesian computations}{in a Bayesian framework as reliability-weighted opinion pooling}.
\replaced[id=JJ]{We demonstrated that the dynamics of conductance-based neuron models naturally computes posterior distributions from Gaussian likelihood functions and prior represented in dendritic compartments}{In this framework, neurons encode posterior distributions via the mean and variance of their somatic membrane potential}.
We derived \replaced[id=JJ]{somatic}{the} membrane dynamics from stochastic gradient ascent on this posterior distribution, and synaptic plasticity from matching the posterior to a \deleted[id=JJ]{somatic }target distribution.
Our plasticity rule naturally accommodates the relative reliabilities of different pathways by scaling up the relative weights of \replaced[id=JJ]{reliable inputs, i.e., those}{those inputs} that have a high correlation to target potentials\added[id=JJ]{ for given presynaptic activities}.
The targets may themselves be formed by peri-somatic input from other modalities, or by more informed predictive input from other cortical areas.
We demonstrated successful learning in a multisensory integration task in which modalities were different in their reliability.


Cortical and hippocampal pyramidal neurons have also been described to be driven by two classes of inputs, with general `top-down' input on apical dendrites that predicts the `bottom-up' input on basal dendrites \cite{Larkum1999,Magee2020}.
In this framework, adapting the basal inputs has been conceptualized as ``learning by the dendritic prediction of somatic firing'' \cite{Urbanczik2014,sacramento2018dendritic,haider2021latent}.
\added[id=JJ]{In the broader context of our Bayesian framework, this view suggests that synaptic plasticity tries to match bottom up input to top-down expectations.
Depending on the nature of the top-down input, learning can be thus interpreted as target matching or -- in the absence of targets -- as a regularization of the cortical representation similar to prior matching in variational autoencoders \cite{kingma2013auto}.}

Our supervised learning can be seen within this predictive framework.
A neuron is considered as a nonlinear prediction element, with dendritic input predicting somatic activity.
\replaced[id=JJ]{Extending this predictive view, we argue that dendrites themselves can be seen as performing a dendritic `opinion pooling' \cite{dietrich2010bayesian,dietrich2016probabilistic}, namely forming dendritic opinions on the stimulus feature, weighting them according to their reliability, and predicting the somatic opinion that is imposed by the teacher input.}{Extending this predictive view, we argue that dendrite themselves can be seen as performing a `cognitive operation', namely forming dendritic opinions, weighting them according to their reliability, and predicting the somatic opinion that is imposed by the teacher input.}
Each dendrite receives a subset of the neuron's afferents and forms its own opinion whether a certain feature is likely present in this afferent subset.
While the dendritic opinion is encoded in the effective dendritic reversal potential, the reliability of this opinion is encoded in the total dendritic conductance.
According to the biophysics of neurons, the overall somatic opinion is then formed by the certainty-weighted dendritic opinions, and this is what the somatic output represents.

So far, we have only considered synapses of which the conductance does not depend on the local membrane potential.
Excitatory synapses in pyramidal cells are known to express N-methyl-D-aspartate (NMDA) channels, whose conductance depends on the local potential \cite{MacDonald1982}.
These synapses elicit strong supra-linear responses \cite{Schiller2000} which cause a massive increase of the isolated dendritic conductance and both dendritic and somatic potentials. 
In our current framework, such responses would correspond to a high certainty that a given feature is present in the input targeting the dendritic branch.
Dendritic calcium spikes that originate in the apical dendrites of layer 5 pyramidal neurons \cite{Larkum1999,London2005} may also represent such strong \replaced[id=JJ]{responses}{opinions}.
At the time of the \deleted[id=JJ]{somatic }peak potential, when the derivative\deleted[id=JJ]{ of this potential} vanishes, these strong \replaced[id=JJ]{responses}{opinions} can be pooled with other dendritic \replaced[id=JJ]{potentials}{opinions}.
As a result, the dendritic spikes can then be integrated according to their reliabilities to form the \replaced[id=JJ]{somatic posterior}{overall dendritic opinion weighting}.
\added[id=JJ]{However, these strongly non-linear, recurrent interactions are difficult to fully capture in the current mathematical framework.
An extended model, which could also describe the influence of backpropagation action potentials necessary for learning, is a promising direction to further reduce the gap to biophysical dynamics.}


Bayesian inference has previously been suggested as an operation on the level of a neuronal population in space \cite{knill2004bayesian,ma2006bayesian,Kording2006} or in time \cite{fischer2011owl,orban2016neural,petrovici2016stochastic,dold2019stochasticity}.
In our framework, to read out the reliability of a single neuron\deleted[id=JJ]{'s opinion about the presence of its presynaptic feature}, postsynaptic neurons either have to average across time or across a population of neurons that encode the same feature.
Our single-neuron description of Bayesian inference \added[id=JJ]{may thus be}{is} complementary to \deleted[id=JJ]{these }population-based models.
\added[id=JJ]{A formal demonstration of this complementarity is beyond the scope of the current manuscript.}
Other recent work also considers the neuronal representation and learning of uncertainty.
\replaced[id=JJ]{For example, i}{I}n line with our plasticity rules, natural-gradient-descent learning for spiking neurons \cite{kreutzer2020natural} predicts small learning rates for unreliable afferents.
A different approach to representing and learning uncertainty is centered on synaptic weights rather than membrane potentials and conductances \cite{aitchison2021synaptic}.
In this model, each synapse represents a distribution over synaptic weights and plasticity adapts the parameters of this distribution.
While being a complementary hypothesis, this normative view does not incorporate neuronal membrane dynamics.

Our model makes various experimental predictions. 

(i) Certainty representation within a neuron: in response to individual whisker touches, our model implies that the somatic potential of somatosensory neurons is driven towards a stimulus-specific reversal potential; this is consistent with measurements in mouse barrel cortex (Fig.~\ref{fig:results-reversal-potentials0}).
Moreover, the model also predicts that the variability of \added[id=JJ]{cumulative }PSP amplitudes (jumps in the postsynaptic membrane potential following a whisker touch) depends on the frequency of whisker touches.
For high frequencies, i.e., small inter-stimulus intervals, the total evoked conductance \replaced[id=JJ]{remains large}{stays consistently high} and the somatic potential \replaced[id=JJ]{"sticks" more to the corresponding}{has less time to drift away from the} reversal potential between stimuli.
Thus, the pre-stimulus variability of the somatic potential decreases, which in turn reduces the CV (coefficient of variation) of PSP amplitudes upon stimulation (consistent with experimental data, cf.~Figs 1C \& 6K in \cite{crochet2011synaptic}).
Similarly, we predict a drop in the CV of the PSPs with increased whisker deflection amplitude.
A stronger, more certain stimulus would lead to stronger presynaptic firing; this \added[id=JJ]{consequently }yields a stronger clamping and hence a smaller post-stimulus variability of the somatic potential, thereby reducing the variability of stimulus-induced PSPs.


(ii) Synaptic plasticity for certainty learning: to test whether the mean and variance of the somatic potential can be learned by dendritic input, one may consider extracellular stimulations of mixed excitatory and inhibitory presynaptic afferents of a neuron while clamping the somatic potential to a fluctuating target.
\replaced[id=JJ]{Our}{The} plasticity rule predicts that initially, when the mean of the target distribution is not yet matched, excitatory and inhibitory synaptic strengths move in opposite directions, i.e., one increases, the other decreases, to jointly match the average somatic membrane potential to the target potential (cf.~Fig.~\ref{fig:results-plasticity}b1).
Then, after the match in the mean has been approximately reached, the excitatory and inhibitory strengths covary in order to match the variance of the target distribution.

(iii) Cross-modal suppression: consider a setting similar to \cite{ohshiro2017neural} in which an animal receives mismatched visual and vestibular cues about a quantity of interest (cf.~Fig.~\ref{fig:results-crossmodal-suppression}).
From a normative perspective, making the visual stimulus less reliable should shift weight to the vestibular input.
Accordingly, our framework predicts that the total synaptic weights from the visual modality should become smaller.
This causes visual cues to have a smaller effect on the somatic membrane potential, and thus, over the course of learning, the firing rate of the bimodal condition should become more similar to the tactile-only condition.

In conclusion, we suggest that single cortical neurons are naturally equipped with the `cognitive capability' of Bayes-optimal \replaced[id=JJ]{integration of information}{opinion pooling}.
Moreover, our gradient-based formulation opens a promising avenue to explain the dynamics of hierarchically organized networks of such neurons.
Our framework demonstrates that the conductance-based nature of synaptic coupling may not be an artifact of the biological substrate, but rather enables single neurons to perform efficient probabilistic inference previously thought to be realized only at the circuit level.

\section*{Methods}

\subsection*{Equivalent somato-dendritic circuit}
The excitatory and inhibitory dendritic conductances, $\gE_i$ and $\gI_i$, are driven by the presynaptic firing rates $r(t)$ through synaptic weights $W^\text{E/I}_i$ and have the form $g^\text{E/I}_i(t) = W^\text{E/I}_i \, r(t)$.
For notational simplicity we drop the time argument in the following.
The dynamics of the somatic potential $\us$ and dendritic potentials $\ud_i$ for the $D$ dendrites projecting to the soma read as
\begin{align}
C \, \dotus & = \gprior (\Eprior - \us) + \sum_{i=1}^D \gsdc_i (\ud_i - \us) \label{Seq:duv} \\
\Cd_i \, \dot{u}^\text{d}_i & = \gL_i (\EL - \ud_i) + \gE_i (\EE - \ud_i) + \gI_i (\EI - \ud_i) + \gdsc_i (\us - \ud_i) \;,
\label{eq:Sduv}
\end{align}
where $C$ and $C_d$ are the somatic and dendritic capacitances, $E^{\Lk/\E/\I}$ the reversal potentials for the leak, the excitatory and inhibitory currents, $\gsdc_i$ the transfer conductance from the $i$th dendrite to the soma, and $\gdsc_i$ in the reverse direction.
By $\gprior$ and $\Eprior$ we denote the somatic conductance and its induced reversal potential, which in the absence of synaptic input to the soma becomes the leak conductance and the leak reversal potential.

We assume that $\Cd$s are small, so that dendritic dynamics are much faster than somatic dynamics and can be assumed to be in equilibrium.
We can thus set $\dot{u}^\text{d}_i$ to zero and rearrange Eqn.~\ref{eq:Sduv} to obtain
\begin{align}
  \ud_i  - \us = \frac{\gd_i}{\gd_i + \gdsc_i} (\Ed_i - \us) \;,
  \label{eq:ud-minus-us}
\end{align}
with dendritic reversal potentials $\Ed_i$ given by Eq.~\ref{eq:results-Ed} and $\gd_i = \gE_i + \gI_i + \gL_i$.
Plugging Eqn.~\ref{eq:ud-minus-us} into Eqn.~\ref{Seq:duv} and using the shorthand notation $\alphasd_i = \frac{\gsdc_i}{\gdsc_i + \gd_i}$, we obtain
\begin{align}
  C \dotus & = \gprior (\Eprior - \us) + \sum_{i=1}^D \alphasd_i \gd_i (\Ed_i - \us) \;,
  \label{eq:methods-cus}
\end{align}
compare Eqn.~\ref{eq:results-somatic-dynamics} in the main manuscript.
\replaced[id=JJ]{These dynamics are}{This is} equivalent to gradient descent ($-\partial E / \partial \us$) on the energy function
\begin{align}
  E(\us) = \frac{\gprior}{2} (\Eprior - \us)^2 + \sum_{i=1}^D \frac{\alphasd_i \gd_i}{2} (\Ed_i - \us)^2 \;,
  \label{eq:methods-energy}
\end{align}
which also represents the log-posterior of the somatic potential distribution, as we discuss below.

\subsection*{Bayesian theory of somatic potential dynamics}
\label{sec:methods-bayesian-theory-of-somatic-potential-dynamics}

Above, we have outlined a bottom-up derivation of somatic dynamics from the biophysics of structured neurons.
In the following, we consider a probabilistic view of single neuron computation and demonstrate that this top-down approach yields exactly the same somatic membrane potential dynamics.

The assumption of Gaussian \replaced[id=JJ]{likelihoods and priors}{membrane potential densities throughout} reflects the fact that the summation of many independent synaptic inputs generally yields a normal distribution, according to the central limit theorem and in agreement with experimental data \cite{Petersen2016}.
We thus consider a prior distribution over $\us$ of the form
\begin{equation}
  p(\us|\Eprior, \gprior) = \frac{1}{Z_0} e^{-\frac{\gprior}{2\lambdae} (\Eprior - \us)^2} \;,
  \label{eq:methods-prior-dist0}
\end{equation}
with parameters $\lambdae,\gprior,\Eprior$ and normalization constant $Z_0$.
Similarly, we define the dendritic likelihood for $\us$ as
\begin{equation}
  p(\Ed_i | \us, \gd_i) = \frac{1}{Z^\text{d}_i} e^{-\frac{\alphasd_i\gd_i}{2\lambdae} (\Ed_i - \us)^2} \; ,
  \label{eq:methods-cond-dend-dist0}
\end{equation}
with parameters $\alphasd_i, \Ed_i, \gd_i$.
According to Bayes' rule, the posterior distribution of the somatic membrane potential $\us$ is proportional to the product of the dendritic likelihoods  and the prior.
If we further assume that dendrites are conditionally independent (\deleted[id=JJ]{conditional }independence of dendritic \replaced[id=JJ]{densities}{likelihoods} given the somatic potential), their joint \replaced[id=JJ]{density}{likelihood} $p(\vecEd \,|\, \us, \vecgd)$ factorizes, yielding
\begin{equation}
  p(\us \,|\, \Eprior, \gprior, \vecEd, \vecgd)  \propto  p(\vecEd \,|\, \us, \vecgd ) p(\us | \Eprior, \gprior) = \prod_{i=1}^D p(\Ed_i | \us, \gd_i ) p(\us | \Eprior, \gprior) \;.
  \label{eq:methods-pusP0}
\end{equation}
Plugging in Eqs.~\ref{eq:methods-prior-dist0} and \ref{eq:methods-cond-dend-dist0}, we can derive that the posterior is a Gaussian density over $\us$ with mean
\begin{align}
  \barEs = \frac{\gprior \Eprior + \sum_{i=1}^D \alphasd_i \gd_i \Ed_i}{\gprior + \sum_{i=1}^D \alphasd_i \gd_i}
  \label{eq:Es}
\end{align}
and inverse variance
\begin{align}
  \bargs = \gprior + \sum_{i=1}^D \alphasd_i \gd_i \;.
  \label{eq:gs}
\end{align}
We thus obtain
\begin{equation}
  p(\us | W,r) \equiv p(\us \,|\, \Eprior, \gprior, \vecEd, \vecgd) = \frac{1}{Z} e^{-\frac{\bargs}{2 \lambdae}(\us - \barEs)^2}\,,
\label{eq:methods-ps0}
\end{equation}
with normalization factor $Z = \sqrt{\frac{2\pi \lambdae}{\bargs}}$.
We switched in Eqn.~\ref{eq:methods-ps0} to the conditioning on $W$ and the presynaptic rates $r$ since these uniquely determine the dendritic and somatic conductances ($\vecgd$), and thus also the corresponding reversal potentials ($\vecEd$).
Here, we use the conventional linear relationship $g=Wr$ between conductances and presynaptic rates. For more complex synapses with nonlinear transmission of the type $g=f(w, r)$, where $f$ can be an arbitrary function, our derivation holds similarly, but would yield a modified plasticity rule.

The energy function from Eqn.~\ref{eq:methods-energy} is equivalent to $E(\us) = - \lambdae \log p(\us|W,r) - \lambdae\log Z = \frac{\bargs}{2}(\us - \barEs)^2$.
Since $Z$ is independent of $\us$, the somatic membrane potential dynamics from Eqn.~\ref{eq:methods-cus} minimizes the energy $E$ while maximizing the log-posterior, 
\begin{equation}
C\dotus = -\frac{\partial E}{\partial \us} =  \lambdae \frac{\partial}{\partial \us} \log p(\us|W,r) \,.
\label{eq:methods-ps1}
\end{equation}
In this form, \added[id=JJ]{it becomes obvious that }the somatic potential moves towards the maximum-a-posteriori estimate (MAP) of $\us$\added[id=JJ]{ in the absence of noise}.
The stochastic version of Eqn.~\ref{eq:methods-ps1} with Gaussian additive noise leads to Eqn.~\ref{eq:results-somatic-dynamics} in the Results, \replaced[id=JJ]{this can be loosely interpreted as performing}{effectively implementing} Langevin sampling from the posterior distribution.

\subsection*{Weight dynamics}
\label{sec:methods-weight-dynamics}
The KL between the target distribution $p^*$ and the somatic membrane potential distribution can be written  as
\begin{align}
    KL[ p^*(\us|r) | p(\us | W, r) ] = -S(p^*) - \mathbb{E}_{p^*}\left[ \log p(\us | W, r) \right] \,.
    \label{eq:methods-KL1}
\end{align}
The entropy $S$ of the target distribution $p^*$ is independent of the synaptic weights $W$.
Stochastic gradient descent on the KL divergence therefore leads to a learning rule for excitatory and inhibitory synapses that can be directly derived from Eqn.~\ref{eq:methods-ps0} (see SI):
\begin{equation}
\dot W_i^\text{E/I} \propto \lambdae \frac{\partial}{\partial W_i^\text{E/I}} \log p(\us^* | W, r) = \alphasd_i \left[ (\us^* - \barEs) \left( E^{\E/\I} - \tildeEd_i \right) + \frac{\alphads_i}{2} \left( \frac{\lambdae}{\bargs} - (\us^* - \barEs)^2 \right) \right] \, r \,,
\label{eq:dWd}
\end{equation}
with $\alphasd_i = \frac{\gsdc_i}{\gdsc_i + \gd_i}$,  $\alphads_i = \frac{\gdsc_i}{\gdsc_i + \gd_i}$ and $\tildeEd_i = \alphads_i \barEs + (1-\alphads_i) \Ed_i$, see also Eqn.~\ref{eq:dWd0} in the Results, where we assumed symmetric coupling conductances between dendritic compartments and soma, i.e., $\gsdc_i = \gdsc_i$.

As discussed in the main text, the two terms in the plasticity rule roughly correspond to adapting the mean and variance of the somatic distribution.
However, the second term $\propto \frac{\lambdae}{\bargs} - (\us^* - \barEs)^2$ depends not only on a mismatch in the variance, but also on a mismatch in the mean of the distribution.
To highlight this, we rewrite the sample $\us^*$ as $\us^* = \mu^* + \sigma^* \xi^*$, the target mean plus a sample from $\mathcal{N}(0, 1)$ scaled with the target variance. Plugging this into the plasticity rule, the first term becomes $\propto (\mu^* + \sigma^* \xi^* - \barEs)$, and the second term becomes $\propto \frac{\lambdae}{\bargs} - (\mu^* + \sigma^* \xi^* - \barEs)^2$.
This form shows that only after the somatic reversal matches the target mean, $\barEs = \mu^* $, will the synapses adapt so that in expectation $\frac{\lambdae}{\bargs} - (\sigma^* \xi^*)^2 \approx 0$.
Because the $\xi^*$ are samples from a standard normal distribution, we conclude that after learning, beside $\barEs = \mu^* $, we also have $\frac{\lambdae}{\bargs} = {\sigma^*}^2$, i.e., the total synaptic conductance is inversely proportional to the variance of the target potential distribution.
\deleted[id=JJ]{This latter variance is itself matched by the variance of the potential $\sigma^2$ produced by the dendritic input as learning minimizes the KL divergence from $p$ to $p^*$.}
For a proof that, in addition, the total synaptic strength on each dendritic branch \replaced[id=JJ]{becomes}{is} inversely proportional to the variance in the presynaptic rate, $W^\text{tot} \appropto \frac{1}{\sigma_r^2}$, see SI.

In the absence of a target distribution, the neuron essentially sets its own targets.
On average, weight changes in the absence of a target distribution are hence zero.
Since for conductance-based synapses only non-negative weights are meaningful, we define the minimal synaptic weight as zero.

\subsection*{Linear coordinates for nonlinear processing}
\label{sec:methods-linear-coordinates-for-nonlinear-processing}

\begin{figure}[tbp!]
    \centering
    \includegraphics[width=.3\textwidth]{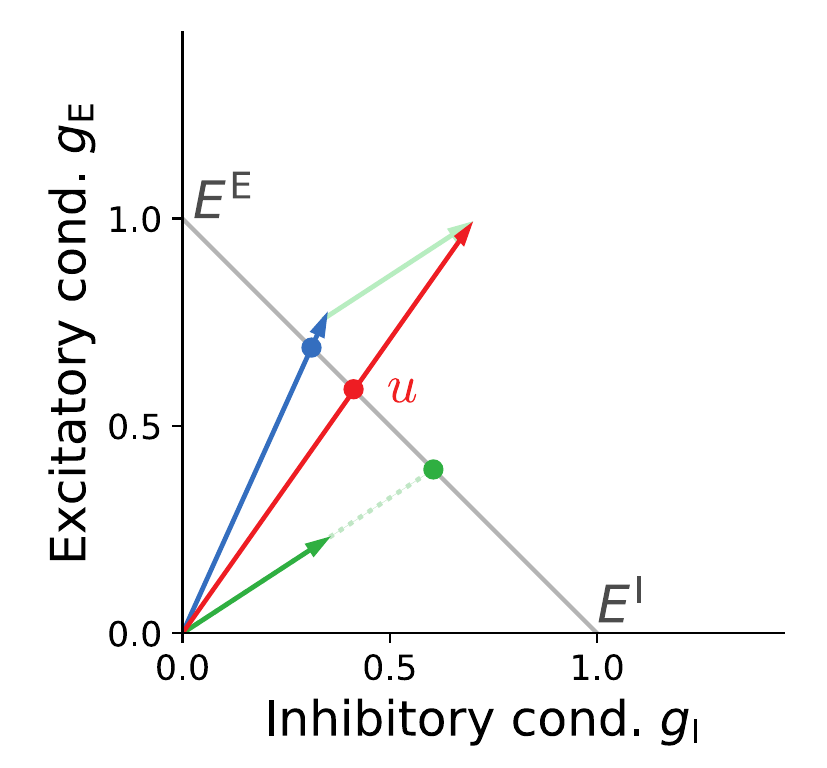}
    \caption{
      {\bf The nonlinear membrane potential and synaptic dynamics expressed in linear conductance coordinates.}
      Dendrites can be represented as vectors defined by their inhibitory and excitatory conductances (blue and green arrows).
      In these coordinates, the soma is itself represented by a vector that is simply the sum of dendritic vectors (red arrow).
      The antidiagonal (gray) spans the range of all possible membrane potentials, from $\EI$ to $\EE$.
      The membrane potential of any given compartment is given by the intersection of its conductance vector with the antidiagonal.
    }\label{fig:SI_Fig2}
\end{figure}

The interplay of conductances and potentials can be visualized in a Cartesian plane spanned by inhibitory and excitatory conductances (Fig.~\ref{fig:SI_Fig2}).
To simplify the picture, we neglect leak conductances and assume strong dendritic couplings $\gsdc, \gdsc$.
The state of a single dendrite is fully determined by its inhibitory and excitatory synaptic conductances and can be represented by a vector $(\gI,\gE)$.
As we assume the prior conductance is zero, the total conductance at the soma is given by the sum of dendritic conductances.
Thus, the soma itself can be represented by a vector that is the sum of the dendritic conductance vectors.
Furthermore, the length of these vectors is proportional to \added[id=JJ]{the magnitude of excitatory and inhibitory conductances and thus }the reliability of the \replaced[id=JJ]{potential}{opinion} encoded by their associated compartments.

This simple, linear construction also allows us to determine the membrane potentials of individual compartments.
For this, we need to construct the antidiagonal segment connecting the points $(1,0)$ and $(0,1)$.
If one identifies the endpoints of this segment with the synaptic reversal potentials, i.e., $\EI \to (1,0)$ and $\EE \to (0,1)$, the antidiagonal can be viewed as a linear map of all possible membrane potentials.
With this construction, the membrane potential of a compartment (dendritic or somatic) is simply given by the intersection of its conductance vector with the antidiagonal.
Formally, this intersection is a nonlinear operation and instantiates a convex combination, the core computation that connects neuronal biophysics to Bayesian inference (Fig.~\ref{fig:results-math-to-neurons}).

This simple construction allows us to easily visualize the effects of synaptic weight changes on the dendritic and somatic \replaced[id=JJ]{membrane potentials}{opinions}.
For example, increasing the inhibitory conductance of a certain compartment will have a twofold effect: its \replaced[id=JJ]{effective reversal potential}{opinion about the presence of its preferred feature} will decrease (the intersection will move towards $\EI$), while simultaneously increasing its reliability (the vector will become longer).

In the following, we give a simple geometric proof that the intersection $u$ of a conductance vector $(\gI, \gE)$ with the antidiagonal indeed represents the correct membrane potential of the compartment.
The coordinates of this intersection are easy to calculate as the solution to the system of equations that define the two lines $x/y = \gI/\gE$ and $y=1-x$, with
\begin{align}
    (x,y) = \left( \frac{\gI}{\gI+\gE}, \frac{\gE}{\gI+\gE} \right) \, .
\end{align}
The ratio of these coordinates is also the ratio of the two resulting segments on the antidiagonal: $(\EE-u)/(u-\EI) = x/y$.
Solving for $u$ yields
\begin{align}
    u = \frac{\gI\EI + \gE\EE}{\gI + \gE} \, ,
\end{align}
which represents the sought convex combination.

\subsection*{Simulation details}
In the following we provide additional detail on simulations.
Numerical values for all parameters can be found in the corresponding tables.

\newlength{\columnwidthleft}
\newlength{\columnwidthmiddle}
\newlength{\columnwidthmiddleconn}

\paragraph{Details to Fig.~\ref{fig:results-reversal-potentials0}}

We consider the trained network from Fig.~\ref{fig:results-orientation-estimation}, but now use a finite somatic capacitance $C$.
The differential equation of the output neurons (Eq.~\ref{eq:results-somatic-dynamics}) is integrated on a time grid of spacing $\Delta t$ with an explicit Runge-Kutta method of order $3(2)$ from SciPy $1.4.1$ \cite{virtanen2020scipy}.
To mimic background noise we generate ``noise'' cues, identical for both modalities, from a normal distribution $\mathcal{N}(\mu_\text{b}, \sigma_\text{b}^2)$ and convert these into rates $r^\text{b}$ via the two populations of feature detectors.
We consider an additional ``signal'' cue, also identical across modalities and trials, which generates additional rates $r'$ via the feature detectors.
The input rate for the output neurons is then computed as $r = \gamma r' + (1 - \gamma) r^\text{b}$, where $\gamma=\gamma^\text{before}$ before stimulus onset and $\gamma=\gamma^\text{after}$ after stimulus onset.
For visualization purposes, we shift the scale of membrane potentials by $-8\text{mV}$ in the figure.

\begin{table*}
  
  \setlength{\columnwidthleft}{0.175\textwidth}
  \setlength{\columnwidthmiddle}{0.25\textwidth}

  \begin{tabularx}{\textwidth}{|p{\columnwidthleft}|p{\columnwidthmiddle}|X|}
    \hline
    \bf Parameter name & \bf Value & \bf Description \\ \hline
    $N_\text{trials}$ & $40$ & number of trials \\
    $\mu^\text{noise}, \sigma^\text{noise}$ & $35^\circ, 15^\circ$ & mean/std.~of noise orientations \\
    $\theta_\text{stimulus}$ & $44^\circ$ & stimulus orientation \\
    $\gamma^\text{before}, \gamma^\text{after}$ & $0.0, 0.88$ & rel.~signal contrast before/after stimulus onset \\
    $dt$ & $0.2\ms$ & integration time step \\
    $T$ & $100\ms$ & simulation duration \\
    $C$ & $50\pF$ & somatic membrane capacitance \\
    $\lambdae$ & $100.0\nS \mV^2$ & neuronal exploration constant \\
    \hline
\end{tabularx}

\caption{Parameters used in Fig.~\ref{fig:results-reversal-potentials0}.
  Remaining parameters defined in Tab.~\ref{tab:supp-nordlie-orientation-estimation}.
}\label{tab:supp-nordlie-reversal-potentials0}

\end{table*}

\paragraph{Details to Fig.~\ref{fig:results-plasticity}}

We consider a neuron following instantaneous versions of Eq.~\ref{eq:results-somatic-dynamics}.
It has $D$ compartments with infinitely strong coupling of the dendritic compartments to the soma $\gdsc, \gsdc \rightarrow \infty$.
In each trial, we sample a ground truth input rate $r \sim \mathcal{N}(\mu_r, \sigma_r^2)$, and from this rate we generate noisy rates $r^\text{V} \sim \mathcal{N}(r, \sigma_\text{V}^2), r^\text{T} \sim \mathcal{N}(r, \sigma_\text{T}^2)$ with modality-specific noise amplitudes $\sigma_\text{V}, \sigma_\text{T}$, respectively.
We avoid non-positive input rates by replacing them with $r_\text{min}$.
We introduce an additional neuron with just a single compartments which generates target membrane potentials $u^*$ from the ground truth input rate $r$ and a random weight matrix.
The second neuron receives the noisy input rates and should learn to mimic the distribution of somatic target potentials by learning synaptic weights via Eq.~\ref{eq:dWd0}.
We train for a certain number of trials $N_\text{trials}$, and for visualization purposes convert trial number into time by defining a trial duration of $\Delta t_\text{trial}$.

\begin{table*}
  
  \setlength{\columnwidthleft}{0.175\textwidth}
  \setlength{\columnwidthmiddle}{0.25\textwidth}

  \begin{tabularx}{\textwidth}{|p{\columnwidthleft}|p{\columnwidthmiddle}|X|}
    \hline
    \bf Parameter name & \bf Value & \bf Description \\ \hline
    $N$ & $1$ & number of neurons \\
    $D$ & $2$ & number of dendritic compartments per neuron \\
    $\gL_0$ & $0.25\nS$ & somatic leak conductance \\
    $\gL_i$ & $0.025\nS$ & dendritic leak conductance \\
    $w_\text{init}^\text{min}, w_\text{init}^\text{max}$ & $0.0\nS\s, 0.019\nS\s$ & min/max value of initial excitatory weights \\
    $w_\text{init}^\text{min}, w_\text{init}^\text{max}$ & $0.0\nS\s, 0.21\nS\s$ & min/max value of initial inhibitory weights \\
    $w_\text{init}^\text{min}, w_\text{init}^\text{max}$ & $0.0\nS\s, 1.07\nS\s$ & min/max value of target excitatory weights \\
    $w_\text{init}^\text{min}, w_\text{init}^\text{max}$ & $0.0\nS\s, 7.0\nS\s$ & min/max value of target inhibitory weights \\
    $\eta$ & $1.25 \cdot 10^{-3}$ & learning rate \\
    $N_\text{trials}$ & $110\,000$ & number of trials \\
    $\Delta t_\text{trial}$ & $10\ms$ & trial duration \\
    $r^*$ & $\mathcal{N}(1.2\pers, 0.5\pers)$ & distribution of input rates \\
    $r_\text{min}$ & $0.001\pers$ & minimal input rate \\
    $\sigma_\text{T}$ & $0.3\pers$ & noise amplitude of tactile modality \\
    $\sigma_\text{V}$ & $0.01875\pers$ & noise amplitude of visual modality \\
    \hline
\end{tabularx}

\caption{Parameters used in Fig.~\ref{fig:results-plasticity}.
  Remaining parameters defined in Tab.~\ref{tab:supp-nordlie-orientation-estimation}.
}\label{tab:supp-nordlie-results-plasticity}

\end{table*}

\paragraph{Details to Fig.~\ref{fig:results-orientation-estimation}}
We consider $N$ output neurons each with $D$ dendritic compartments.
Their dynamics are described by Eq.\ref{eq:results-somatic-dynamics}, but for computational efficiency we consider an instantaneous version of with $C \rightarrow 0$.
We furthermore assume infinitely strong coupling of the dendritic compartments to the soma $\gdsc, \gsdc \rightarrow \infty$.
We use a softplus activation function $\rho(\us) = \log\left( 1 + \exp(\us) \right)$.

We define two homogeneous input populations of $N_\text{T}$ and $N_\text{V}$ feature detectors, respectively, with Gaussian tuning curves.
The output rate of a feature detector in response to a cue with orientation $\theta$ is given by:
\begin{align}
  r(\theta) = r_\text{min} + (r_\text{max} - r_\text{min}) e^{-\frac{\kappa}{2} (\theta - \theta')^2} \;,
\end{align}
with minimal rate $r_\text{min}$, maximal rate $r_\text{max}$, concentration $\kappa$ and preferred orientation $\theta'$.
The preferred orientations $\theta'$ are homogeneously covering the interval $[\theta_\text{min}^\text{fd}, \theta_\text{max}^\text{fd}]$.
All feature detectors from one population project to one dendritic compartment of each output neuron via plastic connections.

Each output neuron additionally receives an input from one presynaptic neuron with fixed rate but plastic weight, allowing it to adjust its prior \replaced[id=JJ]{expectations}{opinions}.

Initial weights are randomly sampled from a zero-mean normal distribution with standard deviation $\sigma_\text{init}^w$.
Training proceeds as follows.
From a ground-truth orientation $\theta^*$ two cues, $\theta_\text{V}$, and $\theta_\text{T}$, are generated by sampling from a Gaussian distribution around a true stimulus value with modality-specific noise amplitudes $\sigma_\text{V}$ and $\sigma_\text{T}$).
The true orientation $\theta^*$ determines the output neurons target rates and hence, via the inverse activation function, target membrane potentials.
The output neuron which should prefer orientations $>45^\circ$ is trained to respond with a rate $r_\text{low}^*$ if $\theta < 45^\circ$ and with a rate $r_\text{high}^*$ if $\theta \ge 45^\circ$.
The other output neuron is trained in the opposite fashion.
Weight changes are following Eq.~\ref{eq:dWd0}.
To speed up training we use batches of size $b$ for $N_\text{train}$ trials with ground truth orientations $\theta^*$ sampled uniformly from $[\theta_\text{min}^\text{train}, \theta_\text{max}^\text{train}]$.
During training, with probability $p_\text{bimodal}$ cues are provided via both modalities, while $1-p_\text{bimodal}$ of all trials are unimodal, i.e., feature detectors of one modality remain silent.

For testing the output neurons are asked to classify $N_\text{test}$ cues uniformly sampled from $[\theta_\text{min}^\text{test}, \theta_\text{max}^\text{test}]$, again perturbed by modality specific noise.
The classification is performed on the combined rate of the two output neurons $r = 0.5\, \left( r_0 + (r_\text{low} + r_\text{high} - r_1) \right)$, where $r_0$ is the rate of the neuron preferring orientations $>45^\circ$ and $r_1$ the rate of the other output neuron.
A ground truth orientation $\theta^*$ is classified as $>=45^\circ$ if $r >= r_\text{low} + 0.5\, \left( r_\text{high} - r_\text{low} \right)$.

\begin{table*}
  
  \setlength{\columnwidthleft}{0.175\textwidth}
  \setlength{\columnwidthmiddle}{0.175\textwidth}

  \begin{tabularx}{\textwidth}{|p{\columnwidthleft}|p{\columnwidthmiddle}|X|}
    \hline
    \bf Parameter name & \bf Value & \bf Description \\ \hline
    $N$ & 2 & number of neurons \\
    $D$ & $3$ & number of dendritic compartments per neuron \\
    $\gL_0$ & $1.0\nS$ & somatic leak conductance \\
    $\gL_i$ & $0.2\nS$ & dendritic leak conductance \\
    $\EE, \EI$ & $0\mV, -85\mV$ & exc.~/inh.~reversal potentials \\
    $\EL$ & $-70\mV$ & leak potential \\
    $\lambdae$ & $1.0\nS\mV^2$ & neuronal exploration constant \\
    $C$ & $\rightarrow 0$ & somatic membrane capacitance \\
    $\gsdc_i,\gdsc_i$ & $\rightarrow \infty$ & somato-dendritic/dendro-somatic coupling conductance \\
    $N_\text{T}, N_\text{V}$ & $70$ & number of feature detectors per modality \\
    $[\theta_\text{min}^\text{fd}, \theta_\text{max}^\text{fd}]$ & $[-315^\circ, 405^\circ]$ & min/max preferred orientations of feature detectors \\
    $\kappa$ & $6.0 \frac{1}{\mydeg^2}$ & concentration (inverse variance) of feature detectors \\
    $r_\text{low}, r_\text{high}$ & $0.75\pers, 16.0\pers$ & min/max rates of feature detectors \\
    $w_\text{init}^\text{min}, w_\text{init}^\text{max}$ & $0.0\nS\s, 0.005\nS\s$ & min/max value of initial excitatory weights \\
    $w_\text{init}^\text{min}, w_\text{init}^\text{max}$ & $0.0\nS\s, 0.024\nS\s$ & min/max value of initial inhibitory weights \\
    $\eta$ & $0.25 \cdot 10^{-4}$ & learning rate \\
    $\sigma_\text{T}$ & $28.5^\circ$ & tactile noise amplitude \\
    $\sigma_\text{V}$ & $13.5^\circ$ & visual noise amplitude \\
    $[\theta_\text{min}^\text{train}, \theta_\text{max}^\text{train}]$ & $[-270^\circ, 360^\circ]$ & min/max of training orientations \\
    $[\theta_\text{min}^\text{test}, \theta_\text{max}^\text{test}]$ & $[-135^\circ, 225^\circ]$ & min/max of testing orientations \\
    $\theta_\text{db}$ & $45^\circ$ & decision boundary \\
    $N_\text{train}$ & $400\,000$ & number of training trials \\
    $N_\text{test}$ & $500\,000$ & number of testing trials \\
    $p_\text{bimodal}$ & $0.9$ & probability of a bimodal trial during training \\
    $b$ & $12$ & batch size \\
    $r_\text{low}^*, r_\text{high}^*$ & $0.75\pers, 16.0\pers$ & low/high target rates \\
    \hline
\end{tabularx}

\caption{Parameters used in Fig.~\ref{fig:results-orientation-estimation}.}\label{tab:supp-nordlie-orientation-estimation}

\end{table*}

\paragraph{Details to Fig.~\ref{fig:results-crossmodal-suppression}}

We consider the trained network from Fig.~\ref{fig:results-orientation-estimation}.
Here we set the cues provided to the feature detectors of the tactile and visual modality to fixed values $\theta_\text{V}, \theta_\text{T}$, respectively.
We introduce two additional parameters, the stimulus intensities $c_\text{V}, c_\text{T}$, which linearly scale the rates of all feature detectors of the respective modality.
For visualization purposes we scale the rate of the output neuron by a factor $r_\text{scale}$.

\begin{table*}
  
  \setlength{\columnwidthleft}{0.175\textwidth}
  \setlength{\columnwidthmiddle}{0.15\textwidth}

  \begin{tabularx}{\textwidth}{|p{\columnwidthleft}|p{\columnwidthmiddle}|X|}
    \hline
    \bf Parameter name & \bf Value & \bf Description \\ \hline
    $\theta_\text{T}$ & $65^\circ$ & orientation of tactile cue \\
    $\theta_\text{V}$ & $50^\circ$ & orientation of visual cue \\
    $c_\text{T}, c_\text{V}$ & $[10^{-3}, 10^2]$ & stimulus contrasts of tactile and visual modality \\
    $r_\text{scale}$ & $2.5$ & output rate scaling factor \\
    \hline
\end{tabularx}

\caption{Parameters used in Fig.~\ref{fig:results-crossmodal-suppression}.
  Remaining parameters defined in Tab.~\ref{tab:supp-nordlie-orientation-estimation}.
}\label{tab:supp-nordlie-crossmodal-suppression}

\end{table*}

\section*{Acknowledgments}
WS thanks M.~Larkum and F.~Helmchen for many inspiring discussions on dendritic processing, and M.~Diamond and N.~Nikbakht for sharing and discussing their data in an early state of this work.
The authors thank all members of the CompNeuro and NeuroTMA groups for valuable discussions.
This work has received funding from the European Union $7$th Framework Programme under grant agreement 604102 (HBP), the Horizon 2020 Framework Programme under grant agreements 720270, 785907 and 945539 (HBP), the Swiss National Science Foundation (SNSF, Sinergia grant CRSII5-180316) and the Manfred St{\"a}rk Foundation.

\section*{Author contributions}
JJ, JS, MAP, WS conceptualized the project; JJ, JS, WW, MAP, WS performed mathematical analyses; JJ, JS, WW performed computational modeling; JJ, MAP, WS wrote the original draft; JJ, WW, MAP, WS wrote the manuscript; all authors reviewed and edited the final manuscript; MAP, WS acquired funding; MAP, WS provided supervision.

\section*{Competing Interests Statement}
The authors declare that they have no competing financial interests.

\printbibliography

\appendix

\newpage
\hrule
\vspace{0.5cm}
{\bf \Huge Supplements}
\vspace{0.5cm}
\hrule

\medskip
\section{Definitions}

The following definitions are used throughout the supplementary material and main manuscript:
\begin{align*}
  \us =& \text{somatic membrane potential} \\
  \lambdae =& \text{neuronal exploration parameter} \\
  W_i^\text{E/I} =& \text{excitatory/inhibitory synaptic weights onto dendrite }i \\
  r =& \text{presynaptic rates} \\
  g_i^\text{L} =& \text{leak conductance on dendrite }i \\
  g_i^\text{E/I} =& W_i^\text{E/I} r, \text{excitatory/inhibitory conductance on dendrite }i \\
  E^\text{L/E/I} =& \text{leak/excitatory/inhibitory reversal potential} \\
  \gprior =& \text{prior conductance} \\
  \Eprior =& \text{prior potential} \\
  \gd_i =& \gE_i + \gI_i + \gL_i\; \text{isolated dendritic conductance} \\
  \Ed_i =& \frac{\gE_i \EE + \gI_i \EI + \gL_i \EL}{\gE_i + \gI_i + \gL_i}\; \text{dendritic reversal potential} \\
  \gsdc_i =& \text{dendro-somatic coupling conductance} \\
  \gdsc_i =& \text{somato-dendritic coupling conductance} \\
  \alphasd_i =& \frac{\gsdc_i}{\gdsc_i + \gd_i}\; \text{dendro-somatic coupling factor} \\
  \alphads_i =& \frac{\gdsc_i}{\gdsc_i + \gd_i}\; \text{somato-dendritic coupling factor}  \\
  \bargs =& \gprior + \sum_{i=1}^D \alphasd_i \gd_i\; \text{total somatic conductance} \\
  \barEs =& \frac{1}{\bargs} \left( \gprior \Eprior + \sum_{i=1}^D \alphasd_i \gd_i \Ed_i \right)\; \text{pooled somatic reversal potential}
\end{align*}

\section{Derivation of the somatic potential distribution}
\label{sec:derivation-somatic-dynamics}

We consider the prior distribution on $\us$ of the form
\begin{equation}
  p(\us|\Eprior, \gprior) = \frac{1}{\Zprior} e^{-\frac{\gprior}{2\lambdae} (\Eprior - \us)^2} \; .
\end{equation}
We consider the dendritic likelihood functions for $\us$:
\begin{equation}
  p(\Ed_i | \us, \gd_i) = \frac{1}{\Zd_i} e^{-\frac{\alphasd_i\gd_i}{2\lambdae} (\Ed_i - \us)^2} \; .
\end{equation}
The posterior over $\us$ is given by
\begin{eqnarray}
  p(\us \,|\, \Eprior, \gprior, \vecEd, \vecgd) & \propto & p(\vecEd \,|\, \us, \vecgd ) p(\us | \Eprior, \gprior) = \prod_{i=1}^D p(\Ed_i | \us, \gd_i ) p(\us | \Eprior, \gprior) \; .
  \label{eq:pusP}
\end{eqnarray}

We first consider the unnormalized posterior, and rewrite it, dropping all terms constant \replaced[id=JJ]{w.r.t.~}{in} $\us$:
\begin{align}
  \prod_{i=1}^D p(\Ed_i | \us, \gd_i ) p(\us | \Eprior, \gprior) \;\propto&\; e^{-\frac{\gprior}{2 \lambdae}(\us - \Eprior)^2} \prod_{i=1}^D  e^{-\frac{\alphasd_i\gd_i}{2 \lambdae}(\us - \Ed_i)^2} \notag \\
  \;\propto&\; e^{-\frac{\gprior + \sum_{i=1}^D \alphasd_i \gd_i}{2 \lambdae} \left( \us^2 - 2\us \frac{\gprior \Eprior + \sum_{i=1}^D \alphasd_i\gd_i \Ed_i}{\gprior + \sum_{i=1}^D \alphasd_i\gd_i} \right)} \notag \\
  \;\propto&\; e^{-\frac{\bargs}{2 \lambdae} (\us - \barEs)^2}
\end{align}
As the density needs to be normalized, we can compute the normalization factor $Z$ directly from this form as a Gaussian integral:
\begin{align}
  Z =& \int d\us\, e^{-\frac{\bargs}{2 \lambdae} (\us - \barEs)^2} \notag \\
  =& \sqrt{\frac{2\pi \lambdae}{\bargs}}
\end{align}
This finally results in the somatic potential distribution:
\begin{align}
  p(\us|W,r) = \frac{1}{Z} e^{-\frac{\bargs}{2 \lambdae}(\us - \barEs)^2}\; .
\end{align}

\section{Derivation of membrane potential dynamics}

We introduce the energy $E$ as the negative logarithm of $p$:
\begin{align}
  E(\us, W, r) := -\log p(\us | W, r) \; .
\end{align}
We obtain potential dynamics from gradient descent on $E$:
\begin{align}
  c_\text{m} \dotus =& -\lambdae \frac{\partial}{\partial \us} E(\us, W, r) \notag \\
  =& \lambdae \frac{\partial}{\partial \us} \log p(\us| W,r) \notag \\
  =& \lambdae \frac{\partial}{\partial \us} \left( -\frac{\bargs}{2 \lambdae}(\us - \barEs)^2 + \frac{1}{2} \log \frac{\bargs}{2\pi \lambdae} \right) \notag \\
  =& \bargs (\barEs - \us) \notag \\
  =& \gprior (\Eprior - \us) + \sum_{i=1}^D \alphasd_i \left( \gL_i (\EL - \us) + \gE_i (\EE - \us) + \gI_i (\EI - \us) \right) \; .
\end{align}

\section{Derivation of weight dynamics}

We want to obtain weight dynamics that approximate gradient descent on the KL:
\begin{align}
  -\lambdae \frac{\partial}{\partial W_i^\text{E/I}} \mathbb{E}_r \left[ \text{KL}(p^*(\us|r) \| p(\us|W,r)) \right]
\end{align}

We first rewrite the KL:
\begin{align*}
  \text{KL}(p^*(\us|r) \| p(\us|W,r)) =& \int d\us\, p^*(\us|r) \log \frac{p^*(\us|r)}{p(\us|W,r)} \\
  =& \int d\us\, p^*(\us|r) \log p^*(\us|r) - \int d\us\, p^*(\us|r) \log p(\us|W,r) \\
  =& -S(p^*(\us|r)) - \mathbb{E}_{\us} \left[ \log p(\us|W,r) \right]
\end{align*}

Here, we can drop the first term as it does not depend on $W$.
We perform stochastic gradient descent in $r$ and $\us$, i.e., we drop the averages and use single samples $r \sim p^*(r), u^* \sim p^*(\us|r)$:
\begin{align}
  \lambdae \frac{\partial}{\partial W_i^\text{E/I}} \mathbb{E}_r \left[ \mathbb{E}_{\us} \left[ \log p(\us|W,r) \right] \right] =& \lambdae \frac{\partial}{\partial W_i^\text{E/I}} \int dr\, p^*(r) \int d\us\, p^*(\us|r) \log p(\us|W,r) \notag \\
  \approx& \lambdae \frac{\partial}{\partial W_i^\text{E/I}} \log p(u^*|W,r) \; ,
\end{align}
where in the last step we plugged in the empirical distribution for $p^*(r)p^*(\us|r)$ consisting of Dirac-delta functions centered on the data points $(r, u^*)$.
We set
\begin{align}
  \dot W_i^\text{E/I} = \eta \lambdae \frac{\partial}{\partial W_i^\text{E/I}} \log p(u^*|W,r)
\end{align}
with some fixed learning rate $\eta$.

We compute the derivative:
\begin{align}
  \lambdae \frac{\partial}{\partial W_i^\text{E/I}} \log p(\us|W,r) =& \lambdae \frac{\partial}{\partial W_i^\text{E/I}} \left( -\frac{\bargs}{2 \lambdae}(\us - \barEs)^2 + \frac{1}{2} \log \frac{\bargs}{2\pi \lambdae} \right) \notag \\
  =& -\frac{1}{2} \frac{\partial \bargs}{\partial W_i^\text{E/I}} (\us - \barEs)^2 - \frac{\bargs}{2} \frac{\partial}{\partial W_i^\text{E/I}} (\us - \barEs)^2 + \frac{\lambdae}{2} \frac{\partial}{\partial W_i^\text{E/I}} \log \bargs
\end{align}

We compute the derivative:
\begin{align}
  \frac{\partial \bargs}{\partial W_i^\text{E/I}} =& \frac{\partial}{\partial W_i^\text{E/I}} \left( \gprior + \sum_{d=1}^D \frac{\gsdc_i}{\gdsc_i + \gd_i} \gd_i \right) \notag \\
  =& \frac{\partial}{\partial W_i^\text{E/I}} \frac{\gsdc_i}{\gdsc_i + \gd_i} \gd_i \notag \\
  =& \left( \frac{\partial}{\partial W_i^\text{E/I}} \frac{\gsdc_i}{\gdsc_i + \gd_i} \right) \gd_i + \frac{\gsdc_i}{\gdsc_i + \gd_i} \frac{\partial}{\partial W_i^\text{E/I}} \gd_i \notag \\
  =& \left( -\frac{\gsdc_i}{(\gdsc_i + \gd_i)^2} \frac{\partial}{\partial W_i^\text{E/I}} \gd_i \right) \gd_i + \frac{\gsdc_i}{\gdsc_i + \gd_i} \frac{\partial}{\partial W_i^\text{E/I}} \gd_i \notag \\
  =& \left[ \left( -\frac{\gsdc_i}{(\gdsc_i + \gd_i)^2} \right) \gd_i + \frac{\gsdc_i}{\gdsc_i + \gd_i} \right] r \notag \\
  =& \alphasd_i \alphads_i r
\end{align}
with $\alphasd_i := \frac{\gsdc_i}{\gdsc_i + \gd_i}$ and $\alphads_i := \frac{\gdsc_i}{\gdsc_i + \gd_i}$.
Note that for symmetric coupling conductances $\alphasd_i = \alphads_i$.

We compute the derivative:
\begin{align}
  \frac{\partial}{\partial W_i^\text{E/I}} (\us - \barEs)^2 =& -2(\us - \barEs) \frac{\partial}{\partial W_i^\text{E/I}} \barEs \notag \\
  =& -2(\us - \barEs) \frac{\partial}{\partial W_i^\text{E/I}} \left[ \frac{1}{\bargs} \left( \gprior \Eprior + \sum_{d=1}^D \frac{\gsdc_i}{\gdsc_i + \gd_i} \gd_i \Ed_i \right) \right] \notag \\
  =& -2(\us - \barEs) \left( -\frac{1}{\bargs} \barEs \frac{\partial \bargs}{\partial W_i^\text{E/I}} + \frac{1}{\gprior} \frac{\partial}{\partial W_i^\text{E/I}} \left[ \frac{\gsdc_i}{\gdsc_i + \gd_i} \gd_i \Ed_i \right] \right) \notag \\
  =& -2(\us - \barEs) \left( -\frac{1}{\bargs} \barEs \frac{\partial \bargs}{\partial W_i^\text{E/I}} + \frac{1}{\bargs} \left[ \frac{\partial}{\partial W_i^\text{E/I}} \frac{\gsdc_i}{\gdsc_i + \gd_i} \right] \gd_i \Ed_i + \frac{1}{\bargs} \left[ \frac{\gsdc_i}{\gdsc_i + \gd_i} \right] E^\text{E/I} r \right) \notag \\
  =& -2(\us - \barEs) \left( -\frac{1}{\bargs} \barEs \alphasd_i \alphads_i r - \frac{\alphasd_i}{\bargs} \left[ \frac{1}{\gdsc + \gd_i} r \right] \gd_i \Ed_i + \frac{\alphasd_i}{\bargs} E^\text{E/I} r \right) \notag \\
  =& -2(\us - \barEs) \frac{\alphasd_i}{\bargs} \left( -\barEs \alphads_i - \left[ \frac{\gd_i}{\gdsc + \gd_i} \right] \Ed_i + E^\text{E/I} \right) r \notag \\
  =& -2(\us - \barEs) \frac{\alphasd_i}{\bargs} \left( E^\text{E/I} - \left[ \alphads_i \barEs + (1 - \alphads_i) \Ed_i \right] \right) r
\end{align}

We compute the derivative:
\begin{align}
  \frac{\partial}{\partial W_i^\text{E/I}} \log \bargs =& \frac{1}{\bargs} \frac{\partial \bargs}{\partial W_i^\text{E/I}} \notag \\
  =& \frac{1}{\bargs} \alphasd_i \alphads_i r
\end{align}

We now put everything together, yielding:
\begin{align}
  \lambdae \frac{\partial}{\partial W_i^\text{E/I}} \log p(u^*|W,r) =& -\frac{1}{2} \frac{\partial \bargs}{\partial W_i^\text{E/I}} (u^* - \barEs)^2 - \frac{\bargs}{2} \frac{\partial}{\partial W_i^\text{E/I}} (u^* - \barEs)^2 + \frac{\lambdae}{2} \frac{\partial}{\partial W_i^\text{E/I}} \log \bargs \notag \\
  =& -\frac{1}{2} \alphasd_i \alphads_i r (u^* - \barEs)^2 + (u^* - \barEs) \alphasd_i \left( E^\text{E/I} - \left[ \alphads_i \barEs + (1 - \alphads_i) \Ed_i \right] \right) r + \frac{1}{2} \frac{\lambdae}{\bargs} \alphasd_i \alphads_i r \notag \\
  =& \left[ (u^* - \barEs) \left( E^\text{E/I} - \left[ \alphads_i \barEs + (1 - \alphads_i) \Ed_i \right] \right) - \frac{\alphads_i}{2} \left( (u^* - \barEs)^2 - \frac{\lambdae}{\bargs} \right) \right] \alphasd_i r \notag \\
  =& \left[ (u^* - \barEs) \left( E^\text{E/I} - \tildeEd_i \right) - \frac{\alphads_i}{2} \left( (u^* - \barEs)^2 - \frac{\lambdae}{\bargs} \right) \right] \alphasd_i r
\end{align}
where we introduced $\tildeEd_i = \alphads_i \barEs + (1 - \alphads_i) \Ed_i$.

\section{Unreliable dendritic inputs are assigned small synaptic strengths}

Here, we provide a proof that the total synaptic strength on a dendritic branch scales inversely with the presynaptic rate fluctuations. Here we explicitly consider the case of two dendritic branches.

The full loss function for two dendrites, targeted by two presynaptic rate vectors $r_1$ and $r_2$,
\begin{align}
  \mathcal{L}(W) =& \mathbb{E}_{p^*(r)} \left[ \mathbb{E}_{p^*(r_1,r_2|r)} \left( \text{KL}\left[ p^*(u|r) || p(u|r_1,r_2,W) \right] \right) \right] \notag \\
  =& \int dr\, p^*(r) \int dr_1\, dr_2\, p^*(r_1,r_2|r) \text{KL}\left[ p^*(u|r) || p(u|r_1,r_2,W) \right]
\end{align}
can be rewritten as
\begin{align}
  \int dr_1\, dr_2\, \int dr\, p^*(r) p^*(r_1|r) p^*(r_2|r) \int du\, p^*(u|r) \left[ \log p^*(u|r) - \log p(u|r_1,r_2,W) \right] \; ,
\end{align}
where we assumed that the input rates $r_1,r_2$ are conditionally independent given the ground truth rate $r$ ($p^*(r_1,r_2|r) = p^*(r_1|r)p^*(r_2|r)$).
We drop all terms which only depend on $p^*$, as they do not depend on the synaptic weights $W$ on which we will perform gradients descent, thus leaving
\begin{align}
  -\int dr_1\, dr_2\, \int dr\, p^*(r) p^*(r_1|r) p^*(r_2|r) \int du\, p^*(u|r) \log p(u|r_1,r_2) \; .
\end{align}
We rearrange the integrals to
\begin{align}
  -\int dr_1\, dr_2\, \int du\, \log p(u|r_1,r_2) \int dr\, p^*(r) p^*(r_1|r) p^*(r_2|r) p^*(u|r) \; .
\end{align}
We now define $p^*(r), p^*(r_i|r)$: the distribution over ground truth rates $r$ is a Gaussian with arbitrary mean and variance, the distribution over input rates $r_i$ are Gaussians around the ground truth $r$ with ``modality-specific'' variances $\sigma_i^2$
\begin{align}
  p^*(r) :=& \frac{1}{\sqrt{2 \pi \sigma_r^2}} e^{-\frac{1}{2 \sigma_r^{2}}(r - \mu_r)^2} \; ,\\
  p^*(r_i|r) :=& \frac{1}{\sqrt{2 \pi \sigma_i^2}} e^{-\frac{1}{2 \sigma_i^2}(r_i - r)^2} \; .
\end{align}
We can rewrite the product of Gaussians appearing in the loss function in the last integral over $r$ (see also \cite{bromiley2018products})
\begin{align}
  p^*(r) p^*(r_1|r) p^*(r_2|r) =& \frac{1}{\sqrt{2 \pi \sigma_r^2}} e^{-\frac{1}{2 \sigma_r^2}(r - \mu_r)^2} \frac{1}{\sqrt{2 \pi \sigma_1^2}} e^{-\frac{1}{2 \sigma_1^2}(r_1 - r)^2} \frac{1}{\sqrt{2 \pi \sigma_2^2}} e^{-\frac{1}{2 \sigma_2^2}(r_2 - r)^2} \notag \\
  =& C(\mu_r, \sigma_r, r_1, \sigma_1, r_2, \sigma_2) \frac{1}{\sqrt{2 \pi \sigma^2}} e^{-\frac{1}{2 \sigma^2} (r - \mu)^2}
\end{align}
with
\begin{align}
  C(\mu_r, \sigma_r, r_1, \sigma_1, r_2, \sigma_2) :=& \frac{\sqrt{2 \pi \sigma^2}}{\sqrt{2 \pi \sigma_r^2} \sqrt{2 \pi \sigma_1^2} \sqrt{2 \pi \sigma_2^2}} e^{\frac{1}{2}\sigma^2\left( \frac{r_1}{\sigma_1^2} + \frac{r_2}{\sigma_2^2} \right)^2 - \frac{1}{2} \left( \frac{r_1^2}{\sigma_1^2} + \frac{r_2^2}{\sigma_2^2} \right)} \; ,\\
  \label{eq:target_variance}
  \frac{1}{\sigma^2} :=& \frac{1}{\sigma_r^2} + \frac{1}{\sigma_1^2} + \frac{1}{\sigma_2^2} \; ,\\
  \label{eq:target_mean}
  \mu :=& \sigma^2 \left( \frac{\mu_r}{\sigma_r^2} + \frac{r_1}{\sigma_1^2} + \frac{r_2}{\sigma_2^2}\right) \; .
\end{align}
For simplicity we consider a target distribution of the somatic voltage given the ground truth rate $r$ that is delta function
\begin{align}
  p^*(u|r) :=& \delta(u - r) \; .
\end{align}
With this definition we can solve the integral over $r$ in the loss function
\begin{align}
  \int dr\, p^*(r) p^*(r_1|r) p^*(r_2|r) p^*(u|r) =& C(\mu_r, \sigma_r, r_1, \sigma_1, r_2, \sigma_2) \frac{1}{\sqrt{2 \pi \sigma^2}} e^{-\frac{1}{2 \sigma^2} (u - \mu)^2} \; ,
\end{align}
and our loss function thus becomes
\begin{align}
  -\int dr_1\, dr_2\, C(\mu_r, \sigma_r, r_1, \sigma_1, r_2, \sigma_2) \int du\, \underbrace{\frac{1}{\sqrt{2 \pi \sigma^2}} e^{-\frac{1}{2 \sigma^2} (u - \mu)^2}}_{p^*(u|r_1,r_2)} \log p(u|r_1,r_2) \; .
\end{align}
Since for learning we will consider derivatives w.r.t.~synaptic weights $W$, we can add a term independent of synaptic weights ($\int du\, p^*(u|r_1,r_2) \log p^*(u|r_1,r_2)$) to again obtain an objective function involving a KL
\begin{align}
  -\int dr_1\, dr_2\, C(\mu_r, \sigma_r, r_1, \sigma_1, r_2, \sigma_2) \text{KL} \left[ p^*(u|r_1, r_2) || p(u|r_1,r_2) \right] \; .
\end{align}

We want to compare the relative influence of the input noise amplitudes $\sigma_i^2$ on the target distribution with the influence of synaptic weights $W$ on the distribution represented by the neuron.
To achieve this, we consider a Taylor expansion of both $p^*(u|r_1, r_2)$ and $p(u|r_1, r_2)$ around the input rates up to second order and compare coefficients of this expansion.
Synaptic plasticity in our model tries to match these two distribution, hence we assume that it also matches these coefficients by minimizing their KL.
For simplicity, we assume $\lambda_\text{e}=1$ in the following.

We compute the first derivative of $p^*(u|r_1,r_2)$ which under our assumptions takes Gaussian form w.r.t.~$r_1$
\begin{align}
  \frac{\partial}{\partial r_1} p^*(u|r_1,r_2) =& p^*(\cdot|\cdot) \frac{\partial}{\partial r_1} \left( -\frac{1}{2\sigma^2}(u - \mu)^2 \right) \notag \\
  =& p^*(\cdot|\cdot) \left( \frac{1}{\sigma^2}(u - \mu) \sigma^2 \frac{1}{\sigma_1^2} \right) \notag \\
  =& p^*(\cdot|\cdot) \left( \frac{1}{\sigma_1^2}(u - \mu) \right) \; .
  \label{eq:first-order-true}
\end{align}
Next we compute the second derivative
\begin{align}
  \frac{\partial^2}{\partial r_1^2} p^*(u|r_1,r_2) =& \frac{\partial}{\partial r_1} \left( p^*(\cdot|\cdot) \left( \frac{1}{\sigma_1^2}(u - \mu) \right) \right) \notag \\
  =& \left( \frac{\partial}{\partial r_1} p^*(\cdot|\cdot) \right) \frac{1}{\sigma_1^2}(u - \mu) + p^*(\cdot|\cdot) \frac{\partial}{\partial r_1} \left( \frac{1}{\sigma_1^2}(u - \mu) \right) \notag \\
  =& p^*(\cdot|\cdot) \left( \frac{1}{\sigma_1^2}(u - \mu) \right)^2 - p^*(\cdot|\cdot) \frac{\sigma^2}{\sigma_1^4} \; .
\label{eq:second-order-true}
\end{align}

Similarly, we compute the first derivative of $p(u|r_1,r_2)$ (see Eqn.~\ref{eq:methods-ps0})
\begin{align}
  \frac{\partial}{\partial r_1} p(u|r_1,r_2) =& \frac{1}{\sqrt{2\pi}} \left( \frac{\partial}{\partial r_1} \sqrt{g} \right) e^\cdot + p(\cdot|\cdot) \frac{\partial}{\partial r_1} \left( -\frac{g}{2} (u - \mu)^2 \right) \notag \\
  =& \frac{1}{\sqrt{2\pi}} \frac{1}{2\sqrt{g}} (\wE_1 + \wI_1) e^\cdot + p(\cdot|\cdot) \left( -\frac{1}{2} (\wE_1 + \wI_1) (u - \mu)^2 + g (u - \mu) \frac{\partial}{\partial r_1} \mu \right) \notag \\
  =& p(\cdot|\cdot) \left( \frac{1}{2g}(\wE_1 + \wI_1) - \frac{1}{2} (u - \mu)^2 (\wE_1 + \wI_1) + (u - \mu) (\wE_1\EE + \wI_1\EI - \mu (\wE_1 + \wI_1)) \right) \notag \\
  =& p(\cdot|\cdot) \left( \frac{1}{2} \left( \frac{1}{g} - (u - \mu)^2 \right) (\wE_1 + \wI_1) + (u - \mu) (\wE_1\EE + \wI_1\EI - \mu (\wE_1 + \wI_1)) \right) \; .
  \label{eq:first-order-neuron}
\end{align}
Taking the second derivative yields
\begin{align}
  \frac{\partial^2}{\partial r_1^2} p(u|r_1,r_2) =& \left( \frac{\partial}{\partial r_1} p(\cdot|\cdot) \right) \left( \cdot \right) + p(\cdot|\cdot) \left( \frac{\partial}{\partial r_1} \left( \cdot \right) \right) \notag \\
  =& p(\cdot|\cdot) \left( \cdot \right)^2 + p(\cdot|\cdot) \bigg( \frac{1}{2} \left( \left( \frac{\partial}{\partial r_1} \frac{1}{g} \right) - \left( \frac{\partial}{\partial r_1} (u - \mu)^2 \right) \right) (\wE_1 + \wI_1) \notag \\ & \quad \quad \quad + \left( \frac{\partial}{\partial r_1} (u - \mu) \right) (\wE_1\EE + \wI_1\EI - \mu (\wE_1 + \wI_1)) \notag \\ & \quad \quad \quad - (u - \mu) \left( \frac{\partial}{\partial r_1} \mu \right) (\wE_1 + \wI_1) \bigg) \notag \\
  =& p(\cdot|\cdot) \left( \cdot \right)^2 + p(\cdot|\cdot) \bigg( \frac{1}{2} \bigg( -\frac{1}{g^2} (\wE_1 + \wI_1) \notag \\ & \quad \quad \quad \quad \quad + 2 (u - \mu) \frac{1}{g} (\wE_1\EE + \wI_1\EI - \mu (\wE_1 + \wI_1)) \bigg) (\wE_1 + \wI_1) \notag \\ & \quad \quad \quad - \left( \frac{1}{g} (\wE_1\EE + \wI_1\EI - \mu (\wE_1 + \wI_1)) \right) (\wE_1\EE + \wI_1\EI - \mu (\wE_1 + \wI_1)) \notag \\ & \quad \quad \quad - (u - \mu) \frac{1}{g} (\wE_1\EE + \wI_1\EI - \mu (\wE_1 + \wI_1)) (\wE_1 + \wI_1) \bigg) \notag \\
  =& p(\cdot|\cdot) \left( \cdot \right)^2 + p(\cdot|\cdot) \bigg( -\frac{1}{2g^2} (\wE_1 + \wI_1)^2 \notag \\ & \quad \quad \quad + (u - \mu) \frac{1}{g} (\wE_1\EE + \wI_1\EI - \mu (\wE_1 + \wI_1)) (\wE_1 + \wI_1) \notag \\ & \quad \quad \quad - \frac{1}{g} (\wE_1\EE + \wI_1\EI - \mu (\wE_1 + \wI_1))^2 \notag \\ & \quad \quad \quad - (u - \mu) \frac{1}{g} (\wE_1\EE + \wI_1\EI - \mu (\wE_1 + \wI_1)) (\wE_1 + \wI_1) \bigg) \notag \\
  =& p(\cdot|\cdot) \left( \cdot \right)^2 + p(\cdot|\cdot) \bigg( -\frac{1}{2g^2} (\wE_1 + \wI_1)^2 - \frac{1}{g} (\wE_1\EE + \wI_1\EI - \mu (\wE_1 + \wI_1))^2 \bigg) \; .
  \label{eq:second-order-neuron}
\end{align}

Now we compare coefficients of the Taylor expansions in $r_1$ around $r_2$, i.e., we assume that input noise amplitudes are small.
From the zeroth order we obtain
\begin{align}
  \left. p^*(u|r_1, r_2) \right|_{r_1 = r_2} = \left. p(u|r_1, r_2) \right|_{r_1 = r_2} \; .
\end{align}
From the first order (Eqs.\ref{eq:first-order-true} \& \ref{eq:first-order-neuron}) we obtain
\begin{align}
  \left. p^*(u|r_1, r_2) \frac{1}{\sigma_1^2} (u - \mu) \right|_{r_1 = r_2} =& \left. p(u|r_1, r_2) \left( \frac{1}{2} \left( \frac{1}{g} - (u - \mu)^2 \right) + (u - \mu) (E_1 - \mu)) \right) (\wE_1 + \wI_1) \right|_{r_1 = r_2} \notag \\
  \left. \frac{1}{\sigma_1^2} (u - \mu) \right|_{r_1 = r_2} =& \left. \left( \frac{1}{2} \left( \frac{1}{g} - (u - \mu)^2 \right) + (u - \mu) (E_1 - \mu)) \right) (\wE_1 + \wI_1) \right|_{r_1 = r_2} \; .
\end{align}
where we used the result from the zeroth order to cancel $p^*(u|r_1,r_2)$ with $p(u|r_1,r_2)$ and introduced $E_1 := \frac{\wE_1\EE + \wI_1\EI}{\wE_1 + \wI_1}$.
Finally, from the second order (Eqs.\ref{eq:second-order-true} \& \ref{eq:second-order-neuron}) we obtain
\begin{align}
  \left. p^*(\cdot|\cdot) \left( \frac{1}{\sigma_1^2}(u - \mu) \right)^2 - p^*(\cdot|\cdot) \frac{\sigma^2}{\sigma_1^4} \right|_{r_1 = r_2} =& \left. p(\cdot|\cdot) \left( \cdot \right)^2 + p(\cdot|\cdot) \bigg( -\frac{1}{2g^2} - \frac{1}{g} (E_1 - \mu)^2 \bigg) (\wE_1 + \wI_1)^2 \right|_{r_1 = r_2} \notag \\
  \frac{\sigma^2}{\sigma_1^4} =& \left. \bigg( \frac{1}{2g^2} + \frac{1}{g} (E_1 - \mu)^2 \bigg) (\wE_1 + \wI_1)^2 \right|_{r_1 = r_2} \; .
\end{align}
Similarly, we consider an expansion in $r_2$ around $r_1$ to obtain an expression similar to the previous line.
We divide these two equations to obtain
\begin{align}
  \frac{\sigma_2^4}{\sigma_1^4} =& \frac{\left. \bigg( \frac{1}{2g^2} + \frac{1}{g} (E_1 - \mu)^2 \bigg) w_1^2 \right|_{r_1 = r_2}}{\left. \bigg( \frac{1}{2g^2} + \frac{1}{g} (E_2 - \mu)^2 \bigg) w_2^2 \right|_{r_1 = r_2}} \; .
\end{align}
Both dendrites are learning to match the same target potentials, hence, we assume that the ratio of excitation and inhibition is identical for both dendrites and thus $E_1 = E_2$. This corresponds to the general setting where the inputs to the both dendrites are not perfectly correlated, and each dendrite thus learns to match the target potential.
With this, the equation simplifies to (after taking the square root)
\begin{align}
  \frac{\sigma_2^2}{\sigma_1^2} =& \frac{w_1}{w_2} \; .
\end{align}
We thus conclude that synaptic plasticity, i.e., stochastic gradient descent on our loss function, not only allows the neuron to match the target distribution, but that in this process it also aligns synaptic weights such that more reliable inputs receive larger synaptic weights.

\section{Dendritic parameters}

\begin{figure}[tbp]
    \centering
    \includegraphics[width=\textwidth]{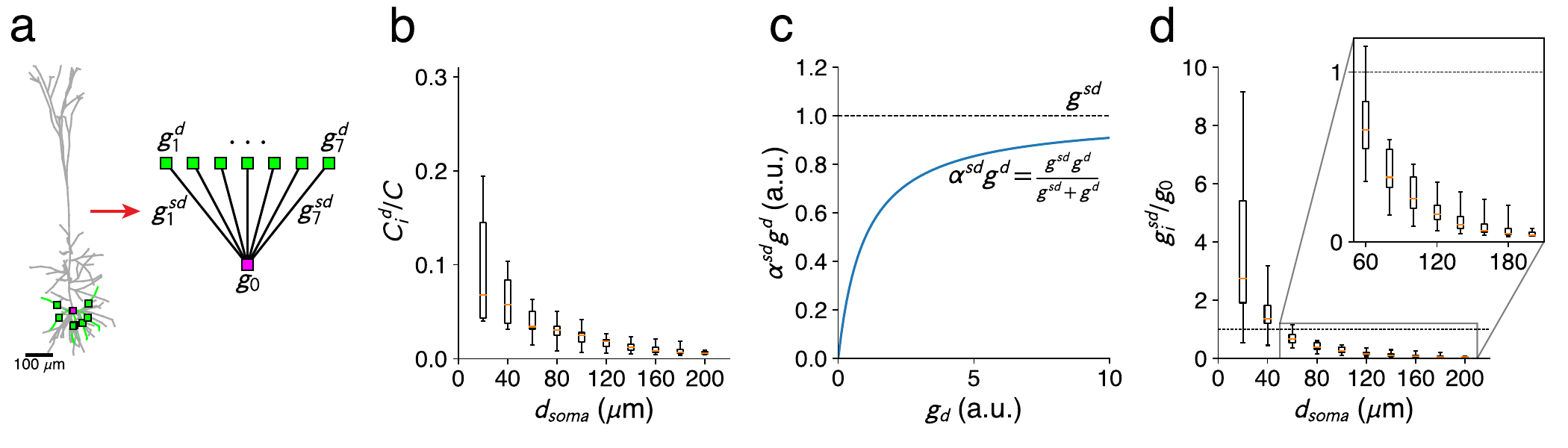}
    \caption{
      {\bf Parameters of the reduced compartmental model as derived from a detailed morphological model.}
      {\bf (a) } A detailed L$5$ Pyramidal cell model (left) is reduced to a configuration with one dendritic compartment on each of seven main basal subtrees (right).
      {\bf (b) } Ratio of dendritic to somatic capacitance, for increasing distances between the dendritic sites and the soma.
      The box indicates the lower and upper quartile values and the orange bar the median. The whiskers indicate the minimal and maximal values. The ratio is always much smaller than one, supporting our approximation of using the instantaneous solution for the dendritic voltage.
      {\bf (c) } Effective dendritic conductance at the soma, $\alphasd_i \, \gd_i$, as a function of the isolated dendritic conductance $\gd_i$.
      This quantity represents the effective reliability of the dendritic \replaced[id=JJ]{potential}{opinion} as read out at the soma. It saturates at the level of the somato-dendritic coupling conductance $\gsdc_i$.
      {\bf (d) } Ratio of the somato-dendritic coupling conductance to the somatic leak conductance for increasing distance between the dendritic site and the soma.
      When this ratio is larger than one, a single branch can overrule the somatic prior.
      Otherwise, multiple branches have to cooperate to overrule the prior.
      The inset shows a magnified version for dendritic sites farther than 50~$\mu$m from the soma.
    }\label{fig:S1}
\end{figure}

Our approach relies on two assumptions with respect to the biophysical model (Eqs.~\ref{Seq:duv}, \ref{eq:Sduv}): the capacitances of the dendritic compartments are small compared to the somatic capacitance and the dendritic conductances $\gd_i$ are able to overrule the somatic prior $g_0$.
A recently developed dendritic simplification framework \cite{Wybo2021} allows us to systematically reduce full biophysical models to obtain the parameters of the reduced compartmental models (Eqs.~\ref{Seq:duv}, \ref{eq:Sduv}) used in this work.
Given a set of dendritic locations on the morphology along the dendritic tree, this approach yields capacitances, leak conductances and coupling conductances for the simplified model that optimally reproduce the dynamics of the full model, at those chosen locations (Fig.~\ref{fig:S1}a).
This, in turns, allows us to assert the validity of the aforementioned assumptions.

We use a detailed biophysical model of an L5 pyramidal cell \cite{Hay2011}.
Without synaptic input, the ion channels in this model collectively determine the cell's prior, encoded in the resting membrane potential and the total conductance at rest. Per dendritic segment, we aggregate these conductance contributions into a single, prior conductance. Formally, this conductance is a passive leak, and the resulting model is a passive model with the same prior (and morphology) as the detailed model.

Then, we choose dendritic sites that allow us to test the validity of our assumptions.
The morphology has seven basal dendritic subtrees with branches of at least $200\mu\text{m}$.
In each subtree, we select one such branch (green in Fig.~\ref{fig:S1}a), and place a single dendritic location on each of those branches at a given distance from the soma.
We increase the distance between soma and dendritic sites in increments of 20~$\mu$m and derive a reduced compartmental model for each configuration (Fig.~\ref{fig:S1}a).
We then compare the ratios of dendritic capacitance $\Cd_i$ and somatic capacitance $C$ for the seven compartments $i \in \{1,\hdots,7\}$.
We find that these ratio are much smaller than one, no matter the distance from the soma (Fig.~\ref{fig:S1}b).

Then, we asses the theoretical maximum degree to which synapses placed at the dendritic sites under investigation can contribute to overruling the somatic prior.
The effective dendritic conductance of compartment $i$, measured at the soma, is given by $\alphasd_i \, \gd_i$ (Eqn.~\ref{eq:methods-cus}).
This function has an asymptotic maximum at the dendro-somatic coupling conductance $\gsdc_i$ (Fig.~\ref{fig:S1}c).
In consequence, $\gsdc_i$ is the theoretical maximal conductance that dendritic synapses in compartment $i$ can exert at the soma.
We thus need to compare $\gsdc_i$ with the somatic prior $\gprior$ (Fig.~\ref{fig:S1}d).
For a distance between soma and dendritic site smaller than $\sim 50\mu$m, we find that a single branch can overrule the prior, as the ratio  $\gsdc_i / \gprior$ is typically larger than one.
For larger distances, multiple branches have to collaborate to overrule the prior (Fig.~\ref{fig:S1}D, inset).

\end{document}